%% file: ms.tex
\documentclass[review]{elsarticle}

\usepackage{lineno}
\usepackage{setspace}
\usepackage{natbib}
\modulolinenumbers[200] 

\usepackage[version=4]{mhchem}

\usepackage{amsmath}
\usepackage{upgreek}

\usepackage{threeparttable}

\usepackage{hyperref}

\hypersetup{
  colorlinks   = true, 
  urlcolor     = red, 
  linkcolor    = blue, 
  citecolor    = blue 
}

\usepackage{mathabx}

\usepackage[defaultcolor=red,%
final,
]{changes}
\definechangesauthor[name={Arnaud Salvador}, color=magenta]{AS}
\definechangesauthor[name={Henri Samuel}, color=orange]{HS}

\graphicspath{{./}{FIG/}}

\journal{Icarus}

\biboptions{authoryear}

\begin{document}

\newcommand\ie{\textit{i.e.}, }
\newcommand\eg{\textit{e.g.}, }

\begin{frontmatter}

\title{Convective outgassing efficiency in planetary magma oceans: insights from computational fluid dynamics}


\author[IPGP,NAU,HABLab,LPL]{Arnaud Salvador\corref{cor1}\fnref{fn1}}
\cortext[cor1]{Corresponding author}
\fntext[fn1]{Present address}
\ead{arnaudsalvador@arizona.edu}
\ead[url]{http://orcid.org/0000-0001-8106-6164}

\author[IPGP]{Henri Samuel}

\address[IPGP]{Universit\'{e}  Paris Cité, Institut de physique du globe de Paris, CNRS, F-75005 Paris, France}
\address[NAU]{Department of Astronomy and Planetary Science, Northern Arizona University, Box 6010, Flagstaff, AZ 86011, USA}
\address[HABLab]{Habitability, Atmospheres, and Biosignatures Laboratory, University of Arizona, Tucson, AZ, USA}
\address[LPL]{Lunar and Planetary Laboratory, University of Arizona, Tucson, AZ, USA}

\begin{abstract}
Planetary atmospheres are commonly thought to result from the efficient outgassing of cooling magma oceans. During this stage, vigorous convective motions in the molten interior are believed to rapidly transport the dissolved volatiles to shallow depths where they exsolve and burst at the surface. This assumption of efficient degassing and atmosphere formation has important implications for both the early and long-term planetary evolution, but has never been tested against fluid dynamics considerations.
Yet, during a convective cycle, only a finite fraction of the magma ocean can reach the shallow depths where oversaturated volatiles exsolution can occur, and a large-scale circulation can exist for vigorously convecting fluids in the presence of inertial effects. This can prevent a substantial magma ocean volume from rapidly reaching the planetary surface. 
Therefore, we conducted computational fluid dynamics experiments of vigorous 2D and 3D Rayleigh-Bénard convection at Prandtl number of unity to characterize the ability of the convecting fluid to reach shallow depths at which volatiles are exsolved and extracted to the atmosphere.
We find that the outgassing efficiency is essentially a function of the magnitude of the convective velocities. This allows deriving simple expressions to predict the time evolution of the amount of outgassed volatiles as a function of the magma ocean governing parameters. 

We show that for plausible cases, the time required to exsolve all oversaturated water can exceed the magma ocean lifetime in a given highly vigorous transient stage, leading to incomplete or even negligible outgassing. Furthermore, the planet size and the initial magma ocean water content, through the convective vigor and the exsolution depth, respectively, strongly affect magma oceans degassing efficiency, possibly leading to divergent planetary evolution paths and resulting surface conditions. Overall, despite vigorous convection, for a significant range of parameters, convective degassing appears not as efficient as previously thought.

\end{abstract}

\begin{keyword}
Terrestrial planets\sep Interiors \sep Geophysics \sep Thermal histories \sep Atmospheres, evolution
\MSC[2010] 00-01\sep  99-00
\end{keyword}

\end{frontmatter}

\linenumbers

\include{main}
\include{appendix}

\section*{Acknowledgments}
We thank the editor, Doris Breuer, for her suggestions and for the handling of our manuscript as well as two reviewers for their comments.
This work has been supported by the Deutsche Forschugsgemeinschaft (project SA 2042/3), and by the INSU-CNES Programme National de Plan\'etologie.
Numerical computations were performed on the S-CAPAD/DANTE platform, IPGP, France.
AS gratefully acknowledges support from NASA's Habitable Worlds Program (No.~80NSSC20K0226).

\bibliography{journal,biblio}

\include{highlights}

\listofchanges

\end{document}

%% file: main.tex
\section{Introduction}

By controlling both the surface temperature and pressure, atmospheres play a fundamental role in planetary evolution and habitability (defined as the ability of a planet to sustain liquid water at its surface; \citealp[\eg][]{Kasting1993b}). The formation of atmospheres is thought to have occurred early in planetary history, through the outgassing of the molten interior, during the magma ocean phase. During this period, accretional heating \citep[\eg][]{Safronov1978,Kaula1979,Tonks1993,Nakajima2021}, radiogenic heating by the decay of short-lived elements \citep[\eg][]{Merk2002, Bhatia2021}, and the conversion of gravitational potential energy into heat via viscous dissipation during core formation \citep[\eg][]{Sasaki1986,Samuel.etal2010}, likely lead rocky bodies to experience at least one episode of global mantle melting \citep{Wood1970,Stevenson81,Elkins-Tanton2012}.


\begin{figure}
\centering
\includegraphics[width=1.\textwidth, height=1.\textheight, keepaspectratio]{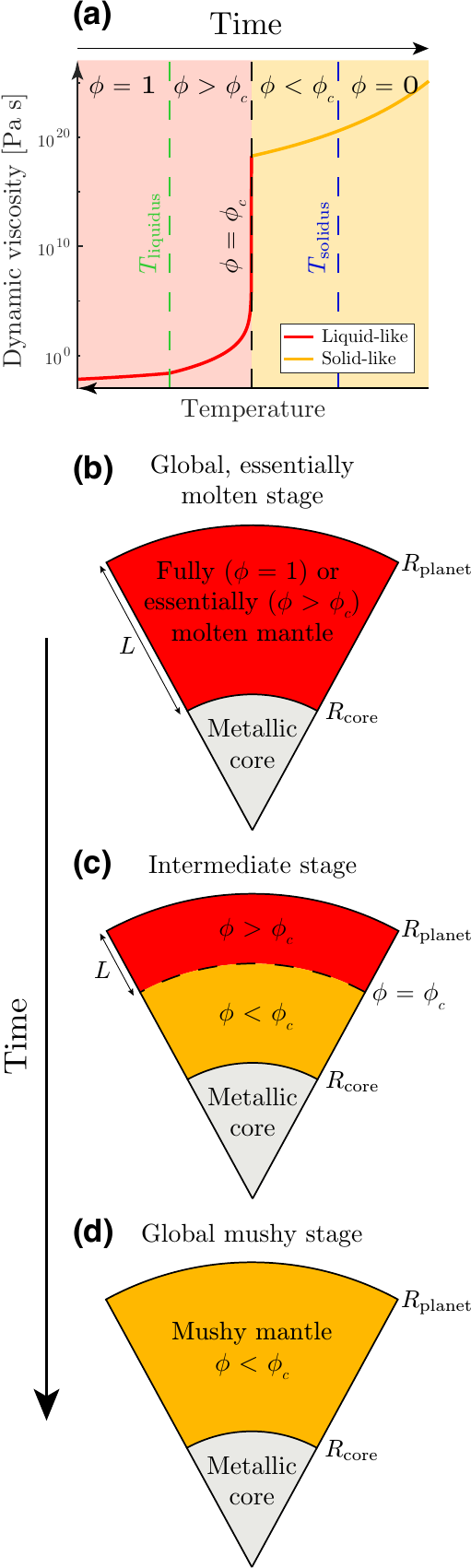}
\caption{(a) Viscosity of the silicate magma ocean as a function of its evolving temperature/crystal fraction \citep[modified from][]{Lejeune&Richet95, Salvador2017}. (b-d) Schematic representation of the succession of three main stages during the progressive solidification of the magma ocean, from (a) a fully ($\phi=1$)/essentially ($\phi_c<\phi<1$) molten stage of global extent, to (b) a mixed essentially molten upper unit coexisting with a lower mushy unit, to (c) a global mushy stage. Red and yellow areas correspond to regions whose rheologies are dominated by the melt ($\phi > \phi_c$) and by the crystals ($\phi < \phi_c$), respectively. The rheological transition and associated viscosity jump occur at the critical liquid fraction $\phi_c$ (black dashed lines in (a) and (c)). During the magma ocean bottom-up solidification, from top to bottom in (b-d), the convective vigor decreases and the inertial effects increase as the viscosity increases and the temperature decreases.}
\label{fig:MO_stages}
\end{figure}


In the classical scenario for magma ocean progressive cooling and crystallization \citep[\eg][]{Solomatov2015} the magma ocean undergoes different major evolution stages that are essentially governed by the rheology of the melt-crystal mixture and by the relationships between the evolving geotherm and mantle melting curves (Figure~\ref{fig:MO_stages}a), leading to changes in the dynamics and in the thickness $L$ of the fully ($\phi=1$, where $\phi$ is the melt fraction) or essentially ($\phi_c<\phi<1$, where $\phi_c$ is a critical melt fraction at which the rheology of the partially molten mixture becomes dominated by that of the solids) liquid magma ocean layer: (i) a global fully/essentially molten stage (Figure~\ref{fig:MO_stages}b), where the entire mantle thickness is either fully molten (because the temperature is hotter than the liquidus at all magma ocean depths) or is partially molten (crystals  form and accumulate at the base of the mantle where the geotherm intersects the liquidus first) but has a melt fraction that exceeds the critical value $\phi_c \approx 0.4$ at all depths \citep{Lejeune&Richet95,Abe1997}. In this case, the partially molten mixture behaves as a liquid throughout the entire mantle thickness, \ie $L=R_{\rm planet}-R_{\rm core}$.
(ii) The geotherm of the cooling magma ocean reaches a temperature corresponding to the critical melt fraction at which the crystals become interconnected, leading to a rheology of the partially molten mixture dominated by the solids. This transition yields an abrupt increase in viscosity (Figure~\ref{fig:MO_stages}a).
The appearance of this evolving rheological front (where $\phi=\phi_c$) marks the separation between two units that convect on very distinct time scales (Figure~\ref{fig:MO_stages}c): the lower, deep mushy unit characterized by a relatively high crystal fraction and a considerably higher viscosity, and the upper, essentially liquid, unit with relatively low crystal fraction, whose rheology remains dominated by that of the melt ($L<R_{\rm planet}-R_{\rm core}$). The viscosity contrast between these two units imply a considerably higher convective vigor with inertial effects in the upper unit, which is the one of interest here. However, as the rheological front progresses upwards, the thickness $L$ of the upper essentially liquid unit decreases, leading to a progressive decrease of convective vigor.  
(iii) When the rheological front reaches the surface, the entire mantle thickness enters a global mushy stage (Figure~\ref{fig:MO_stages}d) characterized by a relatively high viscosity (dominated by the rheology of the interconnected solids) with negligible inertial effects and convective vigor compared to that of the earliest magma ocean stage \citep{Abe1993c,Lejeune&Richet95}.

The stages described above and depicted in Figure~\ref{fig:MO_stages} assume a bottom-up magma ocean crystallization \citep[\eg][]{Andrault2011} but the reasoning would remain the same if the solidification initiates at the mid-mantle as proposed for example in \citet{Labrosse.etal2007} and focus on the region where the solidification front propagates upward. The distribution of chemical elements would however be considerably affected in the presence of several isolated melt reservoirs.

These major stages are composed of a succession of smaller \textquotedblleft{}transient stages\textquotedblright{}, where the motions of the convecting fluid and its properties gradually evolve from a melt- to a crystal-dominated rheology. The entire magma ocean solidification sequence is thus made out of the succession of all these transient stages. The latter correspond to the time intervals during which magma ocean degassing efficiency is evaluated (analogous to the computational time steps used to discretize the magma ocean evolution in numerical simulations).

The duration of each major stage varies according to several more or less constrained quantities such as the initial thermo-chemical state, the planet size, the proximity to the star, the type of orbited star, or the initial amount of volatiles and their influence on key quantities, such as temperature, density, and viscosity. The first part of the earliest stage (\ie the fully molten stage), is the shortest in duration \citep[10–500 years; \eg][]{Solomatov2000,Lebrun2013,Salvador2017,Nikolaou.etal2019,Bower2019} but the most vigorous one. Despite its short lifetime, this stage remains crucial from the point of view of heat transfer and volatiles distribution because the vigor of convective motions favors efficient mixing throughout the entire molten mantle. Even though non-negligible outgassing can continue after the magma ocean stage \citep{Gillmann2014,Gaillard2014,Tosi.etal2017,Gillmann.etal2020,Ortenzi2020,Gaillard2021}, the absence of large-scale melting at the surface and the presence of a stiff, impermeable, and insulating lid on top of the mantle (stagnant-lid or tectonic plates), or even large amounts of water on top of the solid surface of water worlds, limit direct exchanges between the interior and the atmosphere and considerably reduce the outgassing rate and the percolation of volatiles towards the surface \citep{Kite2009,Noack.etal2017,Dorn2018,Kite2018,Krissansen-Totton2021b}. Consequently, planetary outgassing during the earliest magma ocean stages essentially sets up the initial conditions from which rocky planets and their atmospheres evolve until the present day. These early stages of planetary formation may entirely determine the habitability of rocky planets \citep[\eg][]{Kasting1988,Zahnle2007,Hamano2013,Foley2016,Salvador2017,Driscoll2018,Krissansen-Totton2021,Miyazaki2021,Miyazaki2022}.


\begin{figure}
\centering
\includegraphics[width=1.\textwidth, height=.6\textheight, keepaspectratio]{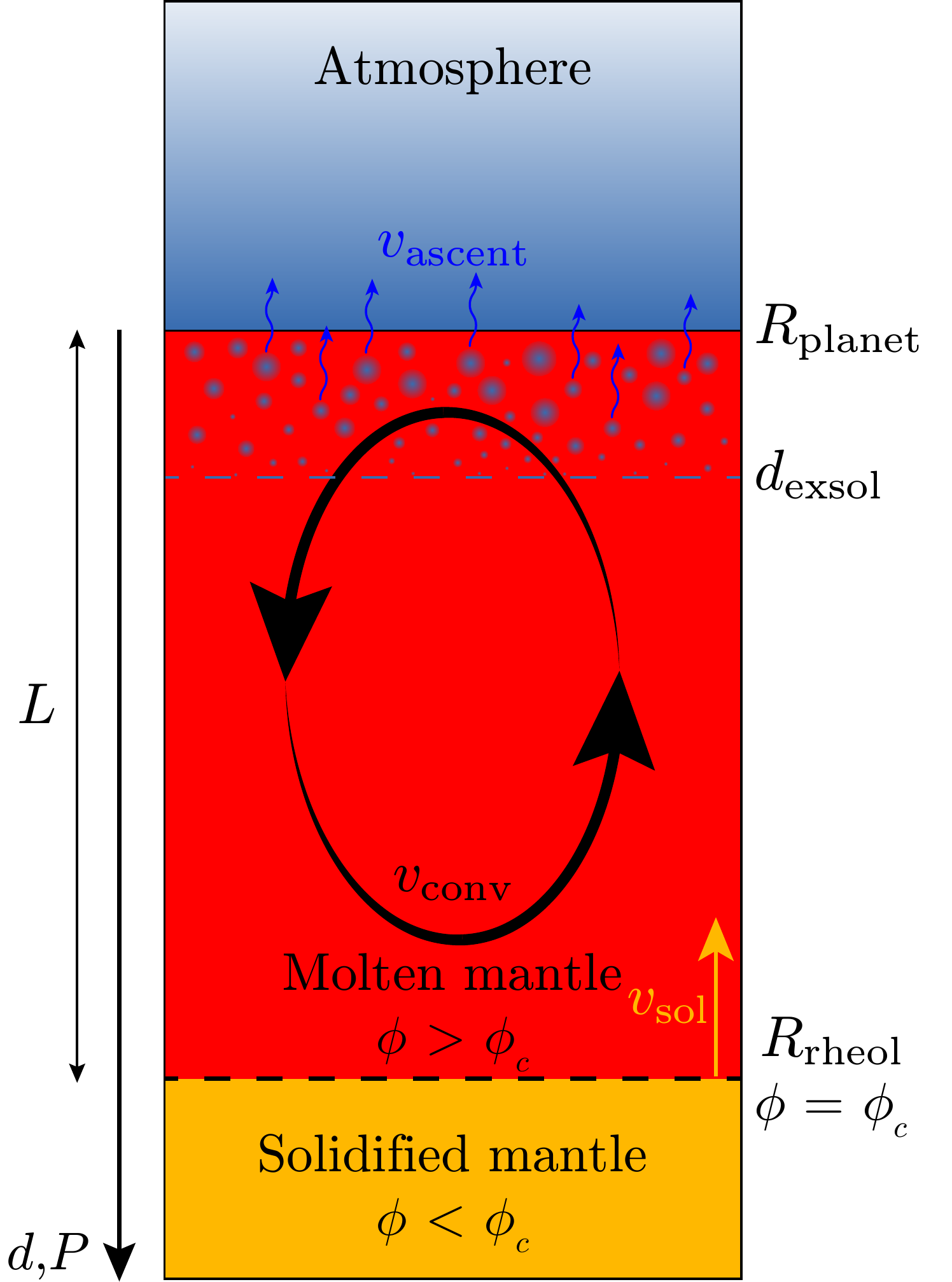}
\caption{Schematic representation of the magma ocean degassing process and associated velocities. $v_{\rm sol}$ is the upward solidification velocity associated with the ascent of the rheological front $R_{\rm rheol}$ (black dashed line) where $\phi = \phi_c$ (assuming that solidification occurs from bottom-up), $v_{\rm conv}$ is the liquid-state convection velocity, $v_{\rm ascent}$ is the ascent velocity of the gas bubbles rising towards the surface. $d_{\rm exsol}$ is the exsolution depth (\ie the dissolved volatile saturation limit) obtained from solubility laws and above which oversaturated volatiles are exsolved out of the melt and start forming gas bubbles. The rheological front separates the low viscosity molten mantle with zero or relatively small crystal fraction (red) and the underlying mushy mantle with solid-dominated rheology (yellow). The thickness $L$ of the vigorously convective magma ocean extends from the surface of the planet down to the rheological transition front.}
\label{fig:degassing_sketch}
\end{figure}


Volatile materials are essentially delivered to a planet during its accretionary phase \citep[\eg][]{Morbidelli2012,Marty2012,OBrien2018,Venturini2020} and are first contained within the molten silicate mantle before being outgassed from the interior to progressively build up an atmosphere according to the following sequence.
(i) Volatile species initially dissolved within the molten mantle are transported by convective motion (black arrows in Figure~\ref{fig:degassing_sketch}). (ii) As the magma travels upward towards the planetary surface, the decrease in the lithostatic pressure allows the volatile species to exsolve out of the melt and form gas bubbles (Figure~\ref{fig:degassing_sketch}).
The depth at which volatiles are exsolved, $d_{\rm exsol}$ (blue dashed line in Figure~\ref{fig:degassing_sketch}), corresponds to the saturation limit of the melt and is given by the solubility of the volatile species (\ie the maximum amount of volatiles that can be dissolved into a silicate melt) \citep{Carroll1994,Papale1997,Berlo2011}. (iii) The gas bubbles subsequently rise up through the upper molten mantle (blue arrows in Figure~\ref{fig:degassing_sketch}) and eventually burst at the surface \citep[\eg][]{Patocka2020}, thus progressively forming the atmosphere \citep[\eg][]{Massol2016,Ikoma2018}. The solubility of the volatile species mainly depends on the pressure and on the volatile concentration but is also a function of melt composition, the type of volatile species mixtures, and temperature \citep[\eg][]{Pan1991,Moore1998,Papale1999,Wallace1999,Keppler2006,Berlo2011}. The outgassing process continues as the melt becomes more and more enriched in incompatible volatile species (\ie the chemical species that preferentially partition into the melt phase) with the gradual solidification of the magma ocean \citep{Elkins-Tanton2008,Lebrun2013}.
The sequence described above implicitly refers to secondary (formed) atmospheres rather than atmospheres resulting from impact degassing \citep[\eg][]{Lange1982, MatsuiAbeNature1986b, Matsui1986b, Zahnle1988, Hashimoto2007, Schaefer2010, Sakuraba2019, Zahnle2020}. Conversely, primary or primordial atmospheres are composed of the solar nebula gases gravitationally captured by the planet \citep{Hayashi1979,Sasaki1990,Ikoma2006,Lammer2014a,Olson2019,Lammer2020}. Primary atmospheres are mainly composed of light elements, \ie H$_2$ and He, that are sensitive to escape processes such as impact erosion or solar extreme-ultraviolet-driven planetary wind, and these atmospheres would have dissipated when the solar nebula disappeared \citep[\eg][]{Zahnle2007,Zahnle2010,Schlichting2015,Schlichting2018,Sakuraba2019,Lammer2020}.

So far, magma ocean degassing, and thus atmosphere formation, has been considered efficient and quasi-instantaneous. This hypothesis is primarily based on one \textit{a priori} robust —yet debatable— assumption that has not yet been fully tested by means of fluid dynamics experiments. It stems from the fact that a magma ocean with a large thickness and very low viscosity convects very vigorously (Figure~\ref{fig:MO_stages}), leading to large convective velocities ($v_{\rm conv}$, black arrows in Figure~\ref{fig:degassing_sketch}) and therefore short transit times. 
The solidification front velocity of the magma ocean ($v_{\rm sol}$, yellow arrow in Figure~\ref{fig:degassing_sketch}) is expected to be significantly smaller than the rapid convective velocities which would then carry volatiles-supersaturated liquids to pressures small enough such that volatiles above melt saturation exsolve into gas bubbles much faster than the magma ocean solidification time (\ie $v_{\rm sol} \ll v_{\rm conv}$). Bubbles would then rise up to the surface fast enough that they would burst into the atmosphere before being carried back down by convective motions (\ie $v_{\rm conv} \ll v_{\rm ascent}$, with $v_{\rm ascent}$ being the bubbles ascent velocity, blue arrows in Figure~\ref{fig:degassing_sketch}) \citep{Elkins-Tanton2008, Ikoma2018}. The assumption of rapid ascent of gas bubbles seems reasonable for magma oceans, as high temperatures, low viscosity magmas allow gas to easily segregate from the melt in the form of small bubbles that can easily merge leading to larger bubbles with a faster ascending speed \citep{Sparks1978,Sparks1994,Lesher2015}. Consequently, most of the magma ocean would have \textquotedblleft{}seen\textquotedblright{} the surface at least once and lost its volatiles in excess of saturation on that occasion. Based on estimates of convective velocities only, complete circulation of the magma ocean up to the surface, and hence exsolution and degassing of all oversaturated volatiles is thought to occur in one to three weeks \citep[\eg][]{Elkins-Tanton2008}.

However, for a vigorously convecting fluid in the presence of inertial effects (the case of a magma ocean), the fluid motions can be organized according to large-scale circulations \citep[\eg][]{Krishnamurti1981,Castaing1989,Lohse&Grossmann92,Siggia1994}, which can isolate a significant part of the magma ocean from the surface, thereby limiting the amount of volatile-supersaturated melt reaching the exsolution depth, and restricting outgassing to the magma ocean fraction reaching shallow depths only.

The implicit assumption of the complete and instantaneous degassing of oversaturated volatile species in a convecting magma ocean has strong implications on the study of planetary atmosphere formation, and on the coupled interior--atmosphere cooling, solidification, and chemical differentiation of magma oceans.
For example, coupled magma ocean--atmosphere numerical models implicitly assume instantaneous degassing of oversaturated volatile species in excess of saturation at all times. This amounts to computing the transfer of volatile species between the planetary interior and the atmosphere via the equilibrium between their volatile content in the gas phase from either side of the planetary surface at each computational time step \citep{Zahnle1988,Elkins-Tanton2008,Hamano2013,Lebrun2013,Hamano2015,Schaefer2016,Salvador2017,Nikolaou.etal2019,Bower2019,Lichtenberg.etal2021,Barth2021,Krissansen-Totton2021}, such that degassing is assumed to be \textquotedblleft{}instantaneous\textquotedblright{} at each modeling time step.
Although the assumption of efficient outgassing in vigorously convecting fluids relying on the large value of the convective velocities may seem sound in the frame of magma ocean modeling, the degassing efficiency in such a context has never been tested against fluid dynamics considerations.
This is particularly important because the early atmospheres and planetary thermo-chemical state exert a considerable influence on the subsequent evolution of rocky bodies until the present day.

Addressing the fundamental question of magma ocean degassing efficiency therefore requires to constrain the amount of melt (containing dissolved volatile species) that reaches the volatile exsolution depth as a function of time, while considering the convective patterns taking place during magma ocean cooling. 
To this end, we conducted computational fluid dynamics experiments of Rayleigh-Bénard convection accounting for inertial effects at a finite Prandtl number value of unity. These experiments can characterize the magma ocean outgassing efficiency as a function of the parameters governing its dynamics.
Then, the resulting description of outgassing efficiency allows one to predict the amount of oversaturated volatiles outgassed during a magma ocean stage as a function of time.

The paper is organized as follows: Section~\ref{sec: methods} provides details of the model and numerical procedure. Section~\ref{sec: results} presents the results of the numerical experiments and relates the observed convective dynamics to the amount of fluid reaching a given depth. This allows estimating the time required for the entire fluid volume to reach arbitrary exsolution depths required for the oversaturated volatile species to fully exsolve and outgas. In Section~\ref{sec: discussion} prior to the conclusion we apply our model predictions to rocky planet magma oceans in a vigorously convecting stage to compare the water depletion timescales and follow the time evolution of magma ocean degassing. To this end, we consider different initial water contents and associated realistic exsolution depths, along with two different planet sizes: an Earth-sized planet and a five Earth mass super-Earth, where we account for both an enhanced and a weak degassing limit scenarios in addition to a more conservative magma ocean parameters setup.

\section{Methods \label{sec: methods}}
To model the dynamics of the magma ocean, we performed numerical experiments using \texttt{StreamV}, a finite-volume code that models the evolution of a convecting fluid by solving the Navier-Stokes and energy equations \citep{Samuel2012a}, under the Boussinesq approximation. These equations written below in non-dimensional form, are the conservation of mass:

\begin{equation}
\nabla\cdot\textbf{u} = 0,
\label{eq:mass}
\end{equation}

\noindent the conservation of momentum:

\begin{equation}
\frac{1}{Pr} \frac{D\textbf{u}}{Dt}= 
- \nabla p
+ \nabla^2\textbf{u}
- Ra ~T ~\textbf{e}_g,
\label{eq:momentum}
\end{equation}

\noindent and  the conservation of internal energy:
\begin{equation}
\frac{D T}{Dt}= \nabla^2 T,
\end{equation}

\noindent where $\textbf{u}$ is the velocity vector, $t$ is time, $p$ is the dynamic pressure, $T$ is the potential temperature, $\textbf{e}_g$ is a vertical unit vector pointing upwards, and $D X/Dt=\partial X /\partial t + \textbf{u} \cdot \nabla X$ is the Lagrangian derivative for the scalar or vector quantity $X$. 

The characteristic scales used to obtain the above non-dimensional equations are the magma ocean thickness $L$ for distance, the diffusion scale $ L^2/\kappa$ for time (with $\kappa$ the thermal diffusivity),  $\kappa/L$ for velocities, the superadiabatic temperature difference between the base and the top of the magma ocean $  \Delta T$ for temperatures, and the viscous scale $\eta \kappa/L^2$ for pressure and stresses.

 As seen in Equation~\eqref{eq:momentum}, the convective dynamics is entirely governed by two dimensionless numbers. 
 One is the Prandtl number, $Pr$, which expresses the ratio of momentum to heat diffusion:

\begin{equation}
Pr=\frac{\nu}{\kappa},
\end{equation}

\noindent where $\nu=\eta/\rho$ is the kinematic viscosity, $\eta$ is the dynamic viscosity, and $\rho$ is the fluid density.
The second governing dimensionless parameter is the thermal Rayleigh number, $Ra$, that compares the thermal buoyancy force that drives convection to the resisting effects of thermal and momentum diffusion, and therefore represents a measure of the convective vigor:
\begin{equation}
Ra=\frac{\alpha g \Delta T L^3}{\kappa \nu},
\end{equation}

\noindent where $g$ is the gravitational acceleration, and $\alpha$ is the thermal expansion coefficient. 
Note that all the physical parameters that enter the $Pr$ and $Ra$ numbers are assumed to be homogeneous, \ie no temperature or pressure dependence is considered.

The Rayleigh number for a terrestrial magma ocean can be as large as $10^{31}$ \citep{Solomatov2000}. Such high value is out of reach for our numerical experiments, because they would imply resolving extremely thin boundary layers that would require a prohibitive associated amount of computational time. 
However, our calculations span Rayleigh numbers from $10^{8}$ to $10^{12}$ and therefore reach the highly turbulent Rayleigh-B\'enard convection regime \citep[\eg][]{Siggia1994, Grossmann2000, Chilla2012}, which allows for reasonable extrapolations of our results to higher Rayleigh number values.
Contrary to the current solid-state planetary mantles \citep{Schubert.etal2001} where inertial effects are negligible compared to viscous effects (\ie the infinite Prandtl number approximation), inertia cannot be neglected in the frame of a vigorously convective magma ocean. We therefore considered a constant Prandtl number $Pr=1$ in our numerical experiments \citep[consistent with experimental constraints on the small viscosity of silicate melts; \eg][]{Urbain1982} and varied systematically the Rayleigh number.

 Most of our experiments were conducted in a 2D ($x,z$) Cartesian domain, with a few cases in 3D Cartesian geometry ($x,y,z$), where the $z$-axis is parallel to the gravity vector and oriented upward. The top and bottom horizontal boundaries of the domain are isothermal, with dimensionless temperatures $T$ set to 0 and 1, respectively, corresponding to Rayleigh-B\'enard convection configuration where the fluid is heated from below and cooled from above, which is relevant for magma oceans.
Along  the vertical sidewalls and at the surface, we considered free-slip boundary conditions, while the bottom boundary is rigid (\ie no-slip velocity boundary conditions), which accounts for the location where the silicates are solid (or above the rheological threshold at which the solid phase dominates the rheology; \eg \citealp{Lejeune&Richet95}).

The domain is discretized using a uniform grid spacing that is fine enough to  properly resolve the thermal and viscous boundary layers and small-scale convective features at all Rayleigh numbers. We used a number of grid cells ranging between 512 and up to 2048 along the vertical direction and between 512 and 1024 grid cells along the horizontal direction. At the lowest $Ra$ values considered here ($10^8$), we used domains of aspect ratio two to avoid the development of a single cell quasi-steady flow, while higher $Ra$ values cases are conducted on square domains in which the convective vigor and smaller scale convective features prevent the development of strictly steady flows, given our boundary conditions. 
Given the smaller $Ra$ values considered in 3D geometry, we used up to 512 cells in the vertical direction, 128 cells along each horizontal directions, and we checked that both viscous and thermal boundary layers were adequately resolved. This choice of smaller horizontal resolution considered is acceptable given the free-slip and thermally insulating boundary conditions applied to the vertical sidewalls.

In 2D geometry, using a pure stream function formulation defined as $(\partial \psi / \partial z, -\partial \psi / \partial x ) = (u_x, u_z)$, the mass and momentum Equations~\eqref{eq:mass} and \eqref{eq:momentum} are recast into a single equation involving a general biharmonic operator \citep[\eg][]{Kupferman2002}.
The algebraic system resulting from the discretization of this Equation is solved using a geometric multigrid method \citep{Brandt82}.
This implementation has been successfully benchmarked against various analytical and numerical solutions \citep{Samuel2012a,Samuel2012b,Samuel2014,Tosi.etal2015,Samuel2018},
including high Reynolds numbers, turbulent flows \citep{Ghia.etal82}.
For 3D domains, {\tt StreamV} relies on primitive variable formulation of the Navier-Stokes equations.
The latter are solved using a second-order in time projection method \citep{Chorin68} on a Staggered grid.
This decomposition yields a Poisson equation for the dynamic pressure, which is solved using a parallel geometric multigrid method \citep{Brandt82} relying on V-cycles with Jacobi relaxations. The gradient of the obtained pressure is then used to correct the velocity to satisfy the incompressibility constraint (Equation~\eqref{eq:mass}).
Even though the benchmark of the approach in the limit of infinite Prandtl number (which relies on another numerical strategy) is documented elsewhere \citep{Samuel&Evonuk2010}, the {\tt StreamV} implementation in the finite Prandtl number case has not been documented in 3D geometry, which is therefore shown in \ref{3Dbenchmark} following a standard convection benchmark \citep{Fusegi.etal91a}.

For each case run, we first let the convecting system evolve until statistical steady-state, which ensures that the system has forgotten its initial condition.
The statistical steady-state is reached when  the top and bottom heat fluxes are statistically equivalent, and when the Root-Mean Squared (RMS) convective velocities oscillate around an asymptotic value. 
Then, we initiate the tracking of the fluid trajectories using passive Lagrangian tracers initially randomly distributed within the domain and passively advected by the flow field. Each tracer represents a fluid parcel. We used a number of passive tracers at least equal to the total number of grid cells of the computational domain, which ensures a uniform density of tracers, and that their vertical distribution allows resolving the boundary layers. 
The purpose of the tracers is to record the shallowest depth reached by each parcel of the convecting fluid, $d_{\rm min}$, during the entire evolution. This quantity is updated every time a Lagrangian tracer reaches a shallower depth, and will then be used to determine the amount of convecting fluid that is able to reach the exsolution depth as a function of time.
After the passive tracers are inserted, we let the system evolve for at least one hundred \textit{transit times} defined as the time spent for the convecting fluid to travel from one horizontal surface to the other.

\section{Results \label{sec: results}}

\subsection{Global properties of the convective flow}

We conducted numerical experiments at $Ra = 10^8$, $10^9$, $10^{10}$, $10^{11}$, and $10^{12}$, and $Pr=1$, leading to turbulent convection relevant to magma ocean dynamics. We describe below the convective dynamics and detail the characteristics of the dynamical regime and associated scalings. Then, in the next Section, we relate our experimental results to the amount of fluid reaching any given depth to make quantitative predictions for the efficiency of magma ocean outgassing.


\begin{figure}
\centering
\includegraphics[width=0.9\textwidth, height=1\textheight, keepaspectratio]{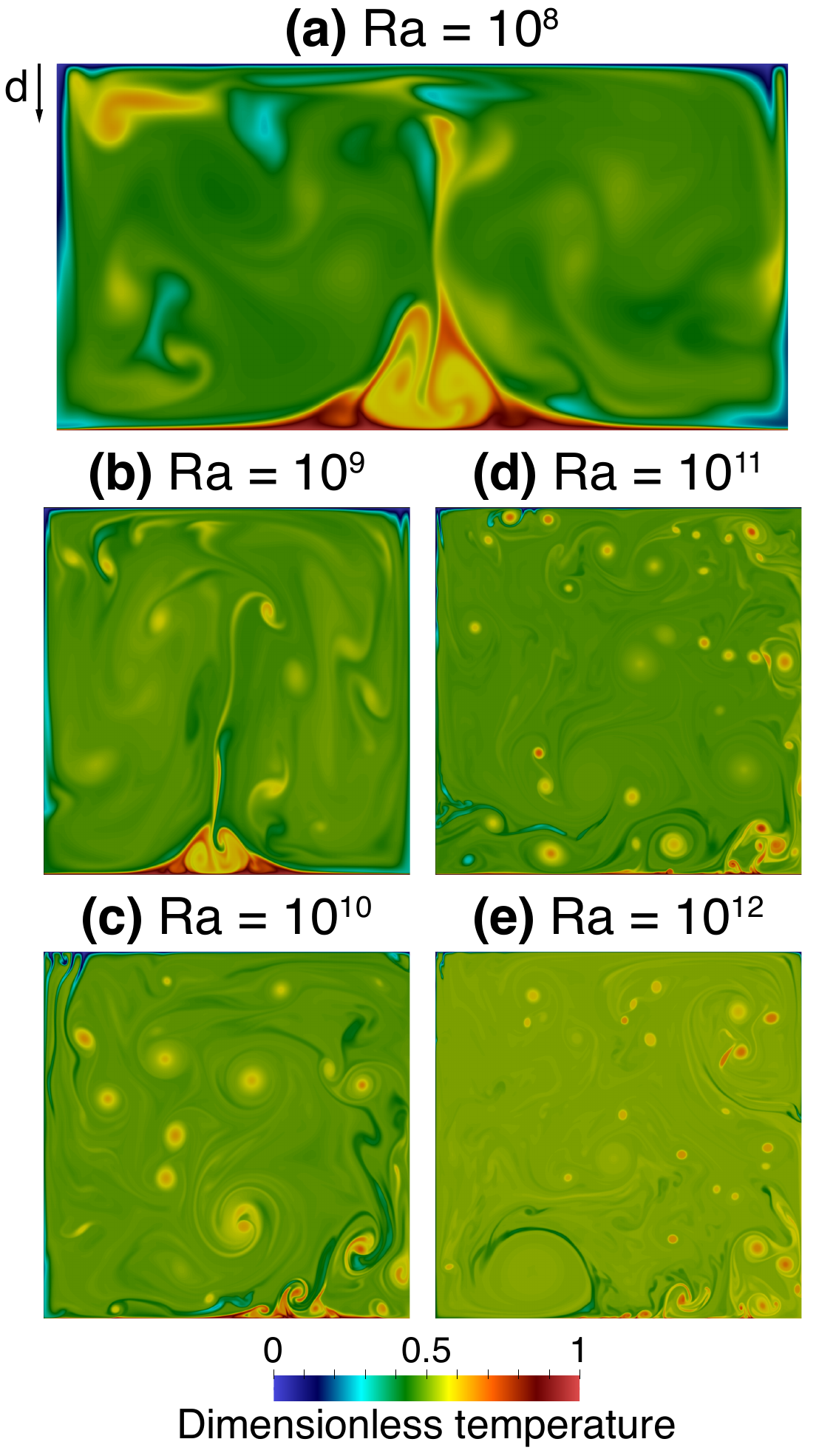}
\caption{Results of the numerical experiments. Temperature fields at statistical steady-state for the different Rayleigh number values considered and for $Pr=1$.}
\label{fig:temperature_field}
\end{figure}


Figure~\ref{fig:temperature_field} shows snapshots of the temperature field for the different Rayleigh numbers at statistical steady-state.
For the range of Rayleigh numbers explored, we observe a large-scale  circulation. The latter is controlled by major ascending hot and major descending cold plumes that define the boundaries of convective cells. The complexity of the hot upwellings and cold downwellings increases with the Rayleigh number through the development of numerous small-scale vortices (Figure \ref{fig:temperature_field}). Therefore, the downwelling and upwelling structures responsible for the large-scale circulation switch from major extended plumes at low $Ra$ numbers (Figures~\ref{fig:temperature_field}a and \ref{fig:temperature_field}b) to small-scale vortices detached from the boundary layers at high $Ra$ values (Figures~\ref{fig:temperature_field}c-e). These small-scale vortices are also present all over the domain at high Rayleigh numbers, as a result of the increased turbulence. 
Such features are consistent with the \textquotedblleft{}hard\textquotedblright{} turbulent convective regime \citep{Heslot1987}, characterized by the existence of a large-scale coherent circulation within the turbulent fluid. This is further confirmed by the exponential probability density function in the temperature distribution observed in our experiments, which is another characteristic of this regime.
Features associated to this convection regime and emphasized here have been extensively described in the literature \citep[\eg][]{Castaing1989, Vincent1999, Vincent2000,Rogers2003}. 

We derived $Nu-Ra$ and $Re-Ra$ scaling relationships, linking respectively the dimensionless Nusselt number, $Nu$, which measures the convective efficiency (transport of heat) relative to diffusion (if $Nu=1$, the heat flux is purely conductive), and the Reynolds number, $Re$, which is the ratio of inertial forces to viscous forces (associated with momentum transport), to the Rayleigh  number.
The Nusselt number is therefore the average dimensionless heat flux taken at either the top or bottom boundary of the domain.


\begin{figure}
\centering
\includegraphics[width=1.\textwidth, height=.7\textheight, keepaspectratio]{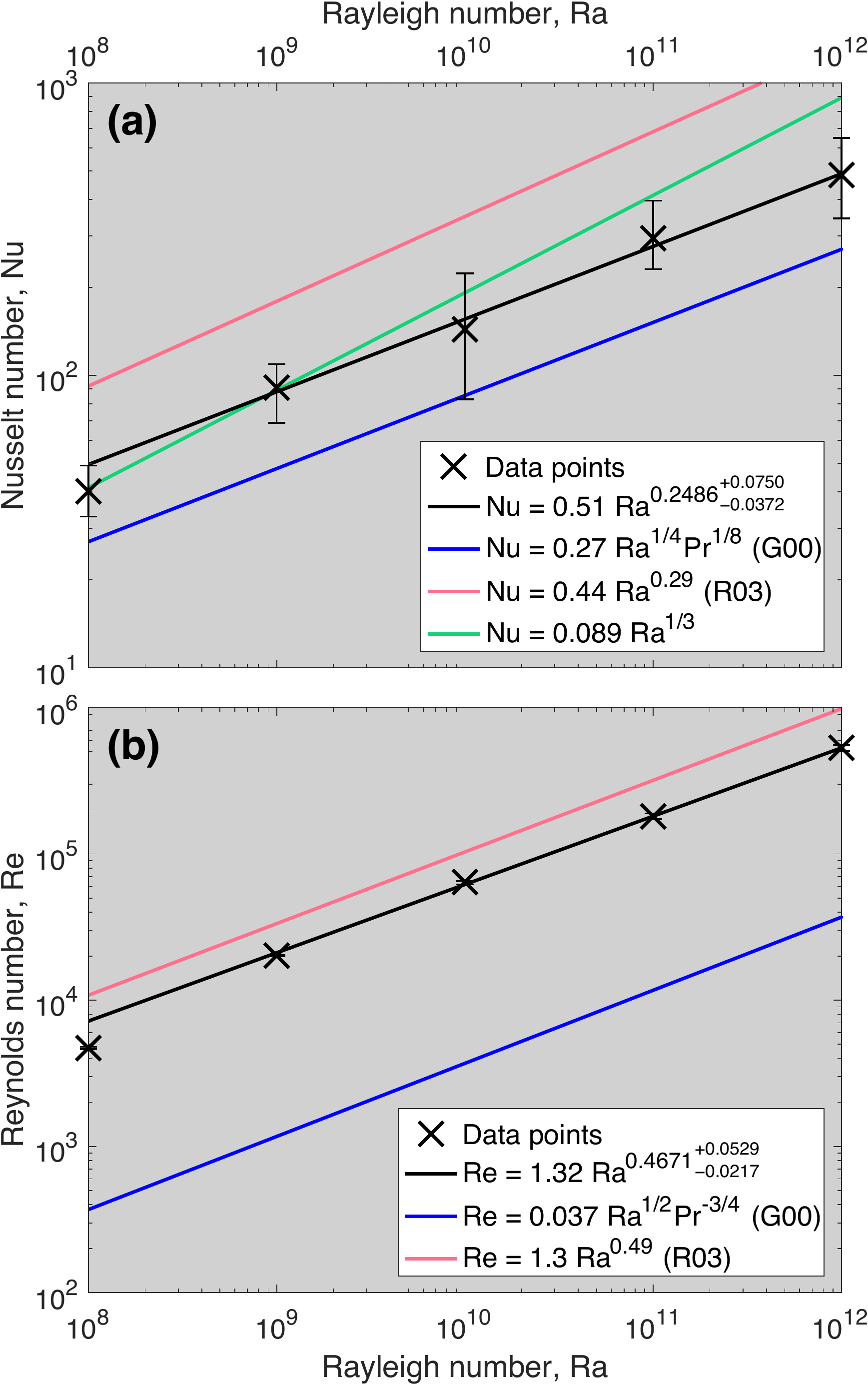}
\caption{(a) Nusselt and (b) Reynolds numbers obtained from our numerical experiments as a function of Rayleigh number (black crosses with uncertainty bars). The corresponding $Nu-Ra$ and $Re-Ra$ scalings derived are shown respectively (black lines). For comparison, scalings obtained from analytic \citep[][referred to as G00]{Grossmann2000} and numerical models \citep[Boussinesq model of][R03]{Rogers2003} are plotted in blue and pink, respectively. The $Nu-Ra$ scaling usually considered in magma ocean studies is shown in green \citep[\eg][]{Siggia1994, Solomatov2015}.}
\label{fig:scalings}
\end{figure}


The dimensionless Reynolds number in our isoviscous numerical experiments with $Pr=1$ corresponds to the magnitude of the convective flow velocity, $U$, where we use the RMS of the magnitude of the velocity vector in the model domain both as a representative measure of $U$ and as a representative measure of the large-scale turbulence in the system \citep[\eg][]{Chilla2012}.
Figures~\ref{fig:scalings}a and \ref{fig:scalings}b show the (statistical) steady-state values of $Nu$ and $Re$, respectively, measured in the experiments as a function of the $Ra$ values investigated.
We observe a power-law increase of $Nu$ and $Re$ with the Rayleigh number:
\begin{equation}
Nu \sim a_{Nu}Ra^{\gamma_{Nu}},
\label{eq:Nu_scaling}
\end{equation}

\begin{equation}
Re \sim a_{Re}Ra^{\gamma_{Re}}.
\label{eq:Re_scaling}
\end{equation}
The dependence between these parameters relies on the still controversial values of both the exponents $\gamma$ and pre-factors $a$ \citep[\eg][]{Malkus1954, Grossmann2000, Chilla2012, Stevens2013, Stevens2018}.
For the calculations conducted here, we found $a_{Nu}=0.51$ and $\gamma_{Nu}=0.25$ (Figure~\ref{fig:scalings}a, black line), using the data points shown in Figure~\ref{fig:scalings}a (black crosses), which correspond to time-averaged values over the 100 convective transits achieved during the second step of our numerical procedure (\ie after statistical steady-state has been reached). Because $Nu$ fluctuates with time around a value at statistical steady-state, the associated range for the power exponent is $\gamma_{Nu}=0.21-0.32$ (data points and corresponding uncertainty bars in Figure~\ref{fig:scalings}a). This is in good agreement with the analytic $\gamma_{Nu}=1/4$ scaling \citep[regime $I_l$ in][blue line in Figure~\ref{fig:scalings}a]{Grossmann2000}, and encompasses the classical experimentally-derived $Nu\sim Ra^{2/7}$ scaling \citep{Castaing1989, Wu1992} as well as outcomes of numerical simulations \citep[Boussinesq model of][pink line in Figure~\ref{fig:scalings}a]{Rogers2003}, that are closer to the canonical $Ra^{1/3}$ scaling \citep[green line in Figure \ref{fig:scalings}a, also corresponding to regime $IV_u$ in][]{Grossmann2000} often considered in magma ocean parameterized convection studies \citep[\eg][]{Lebrun2013, Solomatov2015}. The upper bound $Nu\sim Ra^{1/3}$ scaling would induce significantly larger outgoing heat fluxes at the top of the cell when extrapolating to magma oceans relevant Rayleigh numbers (up to $Ra=10^{31}$), thus implying a more efficient heat transport and cooling. Yet, an exponent lower than $\gamma_{Nu}=1/3$ has been suggested for large-scale circulations \citep{Lebrun2013} and would thus be in agreement with the convective patterns observed here (Figure \ref{fig:temperature_field}).
For the Reynolds-Rayleigh scaling (Equation~\eqref{eq:Re_scaling}), the power-law relationship obtained here takes $a_{Re}=1.32$ and $\gamma_{Re}=0.47$, with the associated uncertainty range $\gamma_{Re}=0.45-0.52$ (Figure~\ref{fig:scalings}b, black line and data points' error bars, respectively). This exponent value is consistent with experimental results of \cite{Castaing1989, Chavanne1997} and with the numerical simulations of \cite{Rogers2003}, where $\gamma_{Re}$ is found to be $\gamma_{Re}=0.49$ for their Boussinesq model (Figure~\ref{fig:scalings}b, pink line). These studies agree with the regime $I_l$ scaling of \cite{Grossmann2000} ($\gamma_{Re}=1/2$, blue line in Figure~\ref{fig:scalings}b), but the  $Re\sim Ra^{4/9}$ scaling of their $IV_u$ regime, associated with higher Rayleigh number values, is even closer to our results.
Note that both $Nu-Ra$ and $Re-Ra$ scalings are sensitive to boundary conditions, aspect ratio, and values of Rayleigh and Prandtl numbers considered \citep[\eg][]{Ahlers2009, Chilla2012,Roche2020}. In our experiments, the velocity and temperature boundary conditions are different from the aforementioned previous studies that consider either rigid, free-slip or periodic boundary conditions for the horizontal and vertical surfaces. Similarly, the aspect ratio and the geometry differ among the different studies. Given these differences and taking into account the uncertainty ranges, we can confidently consider that our scalings are compatible with those obtained in previous works.
The obtained scaling laws will be used to extrapolate our findings to magma oceans relevant $Ra$ numbers and corresponding parameters in Section~\ref{sec: discussion}.

\subsection{Characterization of the fluid ability to reach shallow depths}

As mentioned above, the outgassing efficiency of the magma ocean first relies on the ability of the silicate magma to reach the exsolution depth, regardless of the ability of exsolved volatiles to reach the surface prior to possible re-injection and re-dissolution of oversaturated volatile species at higher lithostatic pressures via downwelling motions.
For this reason, we characterize the ability of vigorously convecting fluid parcels to reach shallow depths as a function of time. This can be seen as the optimistic limiting  case where gas bubbles, once formed at a given exsolution depth, instantaneously reach the surface. In this case, the degassing depth and the exsolution depths are equivalent.


\begin{figure}
\centering
\includegraphics[width=0.7\linewidth]{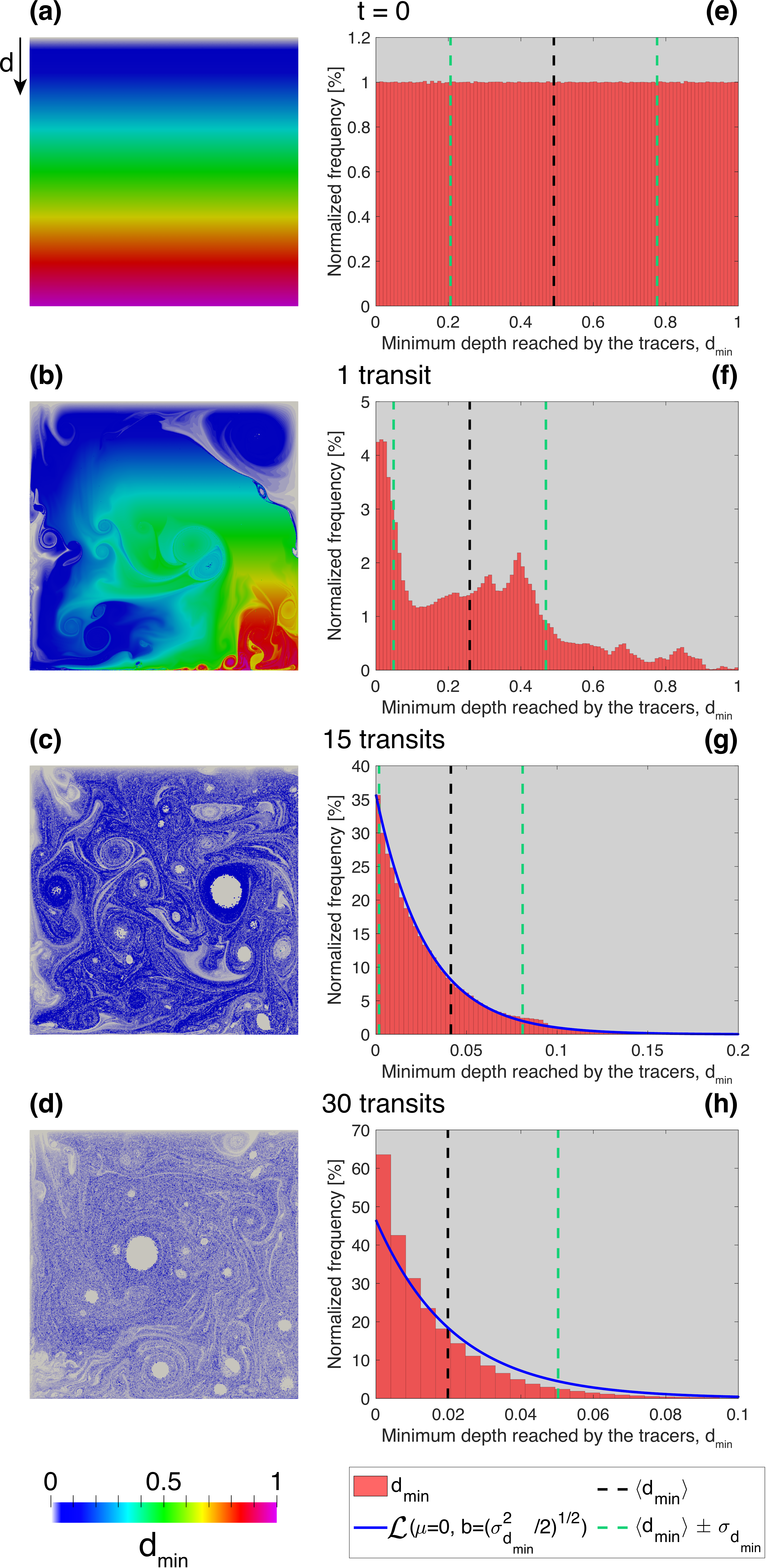}
\caption{Results of the numerical experiments. (a-d) Fields of the dimensionless minimum depth reached by the passive Lagrangian tracers, $d_{\rm min}$, at $t=0$ and for different number of transits for $Ra=10^{12}-Pr=1$. (e-h) Associated histogram distributions. The spatially averaged $d_{\rm min}$ and standard deviation are shown by dashed black and green lines, respectively. The corresponding values for these snapshots are displayed with red crosses in Figures~\ref{fig:min_depth_evolutions}b and \ref{fig:std_min_depth_evolutions}b. Theoretical Laplace probability density function approximating $d_{\rm min}$ distribution is plotted in blue (Equation~\eqref{eq:NLaplace_function}). Note the change of the $x$-axis scale between (e) - (f), (g), and (h).
The steady-state temperature field at $t=0$ corresponding to (a) and (e) is shown in Figure~\ref{fig:temperature_field}e.}
\label{fig:min_depth_snapshots}
\end{figure}


Lagrangian passive tracers track and record the minimum depth reached by any convecting fluid parcel, $d_{\rm min}$, with time. Resuming our simulations from the statistical steady-state stage, Figure~\ref{fig:min_depth_snapshots} shows the time evolution of $d_{\rm min}$ for $Ra=10^{12}$. The corresponding temperature field at the initial time where the tracers are inserted is displayed in Figure~\ref{fig:temperature_field}e. At this time, $d_{\rm min}$ is initialized according to the vertical position of the tracers (Figure~\ref{fig:min_depth_snapshots}a), and therefore ranges from $d_{\rm min}=1$ at the bottom of the domain, to $d_{\rm min}=0$ at the top of the domain.
Since the tracers are uniformly distributed within the domain, the initial $d_{\rm min}$ distribution is uniform (albeit small fluctuations due to the initial random seeding of tracer positions), with a spatially-averaged value $\langle{d_{\rm min}}\rangle=0.5$, corresponding to $0.5 L$, \ie half of the domain height (Figure~\ref{fig:min_depth_snapshots}e).

To represent the time evolution of $d_{\rm min}$, we plot the corresponding field as a function of the number of transits achieved, $N_{\rm transits}$ (Figures~\ref{fig:min_depth_snapshots}b-d), the latter being defined as:

\begin{equation}
N_{\rm transits}(t) = \int_{t=0} ^t U(t)~ {\rm d} t,
\label{eq:N_transits}
\end{equation}
where $U$ is the time-dependent, dimensionless, spatially-averaged magnitude of the flow velocity.
Therefore, the time required to achieve one transit, $N_{\rm transit}=1$, corresponds to the time required for a convection-driven fluid parcel to travel a distance equivalent to that between one horizontal surface to the other. More generally, $N_{\rm transits}$ corresponds to the number of times fluid parcels have traveled on average to distances equal to the thickness of the convective domain.

From the initial time $t=0$ (Figure~\ref{fig:min_depth_snapshots}a), $d_{\rm min}$ only decreases and is updated at each time step for each Lagrangian tracer $i$ as $d^i_{\rm min} (t) ={\rm max}(d^i_{\rm min},1-z^i)$, where $1-z^i(t)$ is the current depth of tracer $i$ (Figures~\ref{fig:min_depth_snapshots}b-d and~\ref{fig:min_depth_snapshots}f-h). Therefore, if a tracer goes downwards, deeper in the domain, the value of its minimum depth reached, $d_{\rm min}$, is not updated and is kept at the smallest depth previously (and ever) reached.
The snapshots of $d_{\rm min}$ fields taken at different times and shown in Figures~\ref{fig:min_depth_snapshots}a-d thus allow to qualitatively estimate the fraction of fluid that reached the surface ($d_{\rm min}=0$, in grey), and that would be outgassed, for different numbers of transit times. Tracking particles this way also gives an idea of the degree of global mixing and homogenization with time.
For example, one can see that vortices are responsible for the formation of isolated blobs of fluid that have been in contact with the surface and that could then be of different composition if, for instance, chemical exchanges would have occurred with the upper, atmospheric layer by the time they were at the surface.

As the number of convective transits achieved increases, $d_{\rm min}$ globally decreases with time, as shown by the increase in grey areas (Figures~\ref{fig:min_depth_snapshots}a-d, for $Ra=10^{12}$). This trend is clearly illustrated by the distribution of the minimum depth reached by the tracers, which becomes more and more asymmetrical and right-skewed with time (Figures~\ref{fig:min_depth_snapshots}e-h).
After one transit time, almost none of the fluid has remained at depths lower than 0.8$L$ (red to magenta areas in Figure~\ref{fig:min_depth_snapshots}b), while a significant portion of the fluid has been entrained upwards, such that $d_{\rm min}$ values associated with  individual tracers have globally decreased, and so did their spatially-averaged value $\langle{d_{\rm min}}\rangle$ (Figure~\ref{fig:min_depth_snapshots}f). 
After 15 transit times, the entire fluid has traveled above the upper fifth of the domain height such that $d_{\rm min}<0.2$ everywhere, as seen in Figures~\ref{fig:min_depth_snapshots}c and \ref{fig:min_depth_snapshots}g. From then on, the distribution of $d_{\rm min}$ is strongly asymmetric, right-skewed, and peaks towards the surface (Figures \ref{fig:min_depth_snapshots}g and \ref{fig:min_depth_snapshots}h).
We further discuss below how such a truncated asymmetric distribution can be described by a Laplace probability density function (blue line in Figures~\ref{fig:min_depth_snapshots}g and \ref{fig:min_depth_snapshots}h).
After 30 transit times, almost all tracers have reached the upper tenth of the domain height ($d_{\rm min}<0.1$, Figures~\ref{fig:min_depth_snapshots}d and \ref{fig:min_depth_snapshots}h). Yet, a non-negligible fraction of the convecting fluid has not reached the surface and additional transits are required for the entire fluid volume to do so, thus demonstrating that in spite of vigorous convection, the ascent of fluid towards shallow depths is neither instantaneous nor negligible in time, contrary to what is commonly assumed in the context of magma ocean degassing studies \citep{Elkins-Tanton2008,Hamano2013,Lebrun2013,Hamano2015,Schaefer2016,Massol2016,Salvador2017,Nikolaou.etal2019,Bower2019,Lichtenberg.etal2021,Barth2021}.

This description of the minimum depth reached by the fluid can be used to estimate magma ocean outgassing efficiency, as previously explained. For instance, if one considers that oversaturated volatile species dissolved within the fluid are exsolved and outgassed as soon as the fluid has reached the surface ($d_{\rm min}=0$), this implies that the corresponding grey areas in Figures~\ref{fig:min_depth_snapshots}a-d have outgassed all the volatile species above saturation. Therefore, even though outgassing increases with the number of transits achieved, more than 30 transit times are required for the entire fluid to reach the surface and to fully outgas oversaturated volatiles.


\begin{figure}
\centering
\includegraphics[width=0.9\textwidth, height=1\textheight, keepaspectratio]{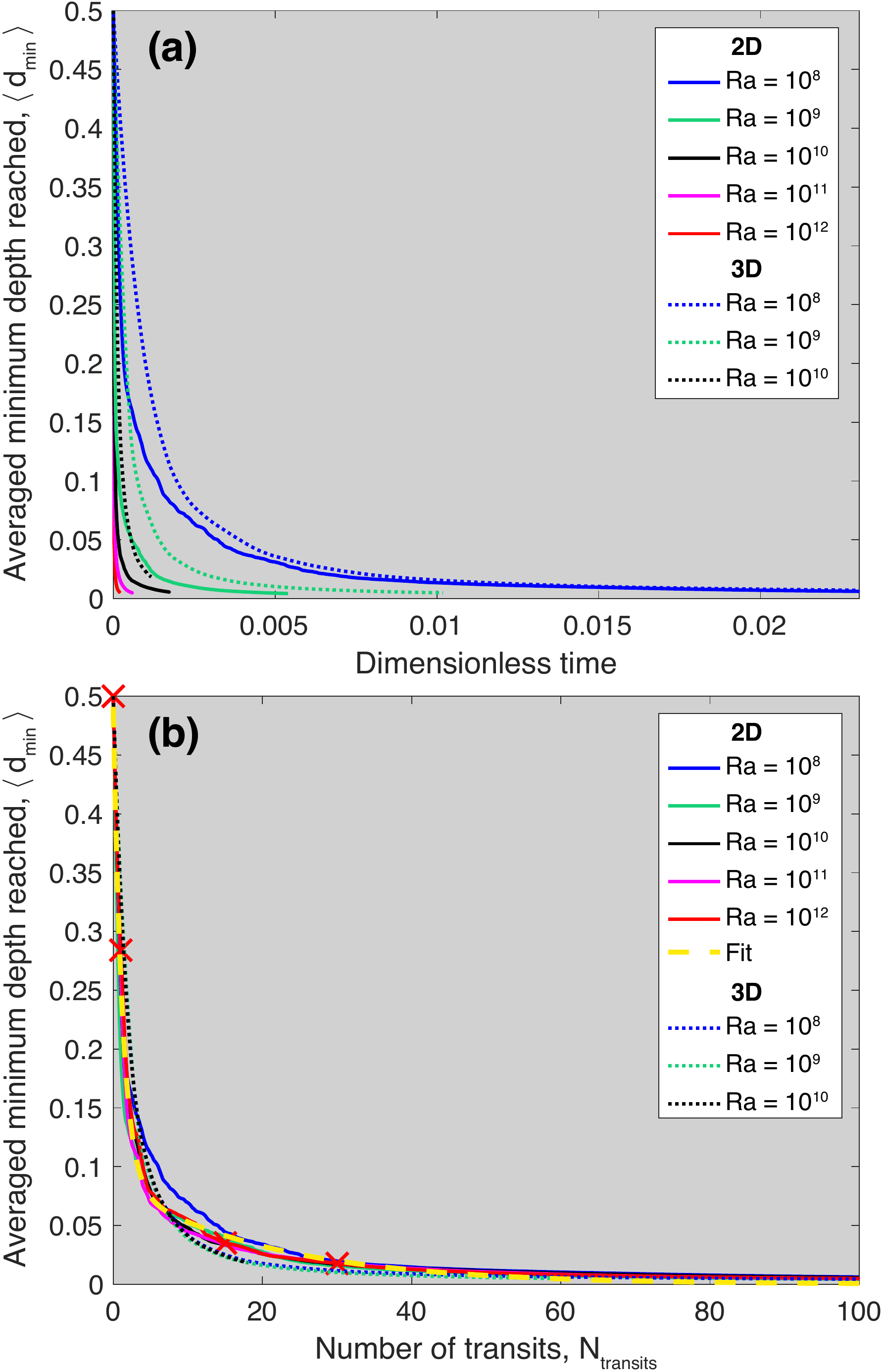}
\caption{Results of the numerical experiments. Spatially averaged minimum depth reached by the tracers $\langle d_{\rm min} \rangle$ as a function of (a) the dimensionless time, and (b) the number of transits achieved for different $Ra$ values. Crosses indicate the values of $\langle d_{\rm min} \rangle$ for the snapshots shown in Figure~\ref{fig:min_depth_snapshots} for $Ra=10^{12}$. The expression of the fit common to all curves (yellow dashed curve) is given in Equation~\eqref{eq:d_min_MEAN}. Results of 3D simulations are shown with the dotted lines.}
\label{fig:min_depth_evolutions}
\end{figure}


To quantify the amount of fluid that reaches a given depth as a function of time, we use the minimum depth reached by the tracers averaged over the entire population of the particles used, $\langle{d_{\rm min}}\rangle$:
\begin{equation}
\langle{d_{\rm min}}\rangle = \frac{\sum^{N_{\rm tracers}}_{i=1} (d^{^i}_{\rm min})}{N_{\rm tracers}},
\end{equation}
where $N_{\rm tracers}$ is the total number of tracers.
Figure~\ref{fig:min_depth_evolutions}a shows the time evolution of $\langle{d_{\rm min}}\rangle$ for the different Rayleigh number values considered in this study. $\langle{d_{\rm min}}\rangle$ appears to strongly depend on the value of the Rayleigh number. The decrease of $\langle{d_{\rm min}}\rangle$ is faster when the Rayleigh number is larger. This is due to the fact that convective velocities, and thus convective efficiency, increase with the Rayleigh number so that the time needed to achieve the same degree of mixing and homogenization is much shorter at higher Rayleigh numbers.
However, using the number of transits (Equation~\eqref{eq:N_transits}) instead of the dimensionless time shows that the ability of the convective flow to bring melt upward is mainly governed by the magnitude of the convective velocities (or the effective Reynolds number). Indeed, by doing so, all the curves obtained for different Rayleigh numbers collapse and can be fitted by a single master curve (Figure~\ref{fig:min_depth_evolutions}b). Importantly, this result implies that efficiency of outgassing mainly shows the same dependence on the convective velocities regardless the value of the Rayleigh number. Note that since convective velocities are related to the value of the Rayleigh number (as seen in Figure~\ref{fig:scalings}b) the convective outgassing efficiency does depend on $Ra$.

We observed two distinct phases for the evolution of $\langle{d_{\rm min}}\rangle$: first, a strong and rapid decrease occurs, during the first two transit times (required for the system to \textquotedblleft{}forget\textquotedblright{} the initial position of the tracers), followed by a more slowly decreasing trend of $\langle{d_{\rm min}}\rangle$ (Figure~\ref{fig:min_depth_evolutions}b).
For example, for $Ra=10^{12}$, within the first transit time, the averaged minimum depth reached by the tracers goes from $\langle{d_{\rm min}}\rangle_{t=0} = 0.5L$ to $\langle{d_{\rm min}}\rangle_{N_{\rm transits}=1} \approx 0.28L$ (first and second red crosses in Figure~\ref{fig:min_depth_evolutions}b, and vertical black dashed lines in Figures~\ref{fig:min_depth_snapshots}e and \ref{fig:min_depth_snapshots}f, respectively), while after 15 convective transits and during the following 15 transit times, the averaged minimum depth reached by the tracers goes from $\langle{d_{\rm min}}\rangle_{N_{\rm transits}=15} \approx 4\times10^{-2}L$ to $\langle{d_{\rm min}}\rangle_{N_{\rm transits}=30} \approx 2\times10^{-2}L$ (third and last red crosses in Figure~\ref{fig:min_depth_evolutions}b, and vertical black dashed lines in Figures~\ref{fig:min_depth_snapshots}g and \ref{fig:min_depth_snapshots}h, respectively).
This implies that for a $L=3000$~km thick magma ocean, and after 30 transit times, the convective fluid would have reached on average a minimum dimensional depth of $\approx 60$~km. These results are in agreement with the qualitative behavior described earlier in Figure~\ref{fig:min_depth_snapshots}: the first few transit times are characterized by a strong mixing and fluid upwelling through the entire domain, while a consequent number of subsequent transits need to be achieved to substantially decrease $\langle{d_{\rm min}}\rangle$, and for the whole fluid to reach the surface or even shallow depths such as $2.5\times10^{-2}L$ (Figure~\ref{fig:min_depth_evolutions}b).

For all Rayleigh numbers we found that $\langle{d_{\rm min}}\rangle$ decreases exponentially as a function of the number of transits and the different curves obtained can be fitted by the following unique expression:
\begin{equation}
\langle{d_{\rm min}}\rangle = 0.4048 \exp[-0.8172 N_{\rm transits}] + 0.08618 \exp[-0.0489 N_{\rm transits}],
\label{eq:d_min_MEAN}
\end{equation}

\noindent which is displayed with the yellow dashed line in Figure~\ref{fig:min_depth_evolutions}b.


\begin{figure}
\centering
\includegraphics[width=0.9\textwidth, height=1\textheight, keepaspectratio]{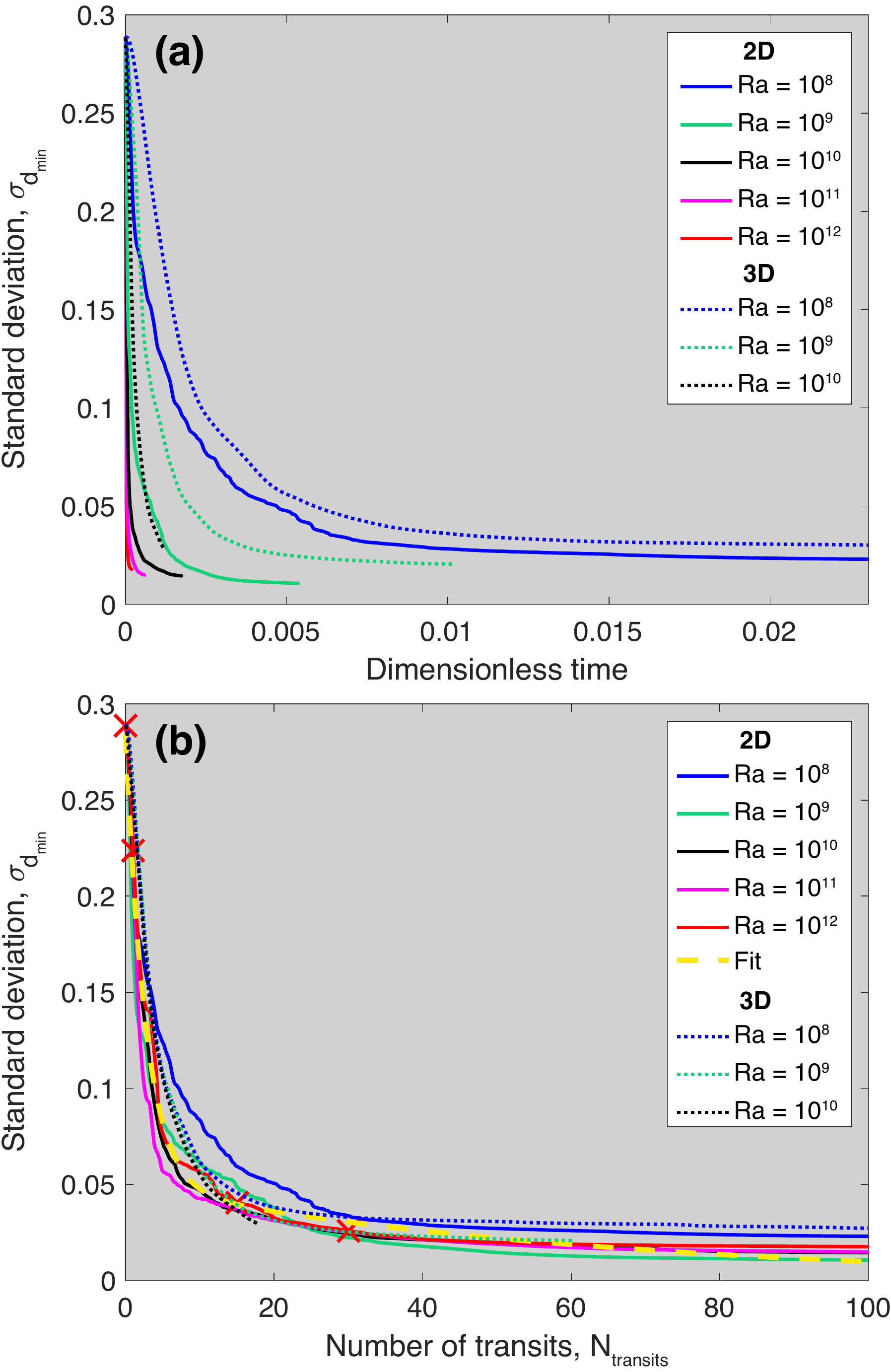}
\caption{Results of the numerical experiments. Standard deviation of the minimum depth reached by the tracers over the entire domain, $\sigma_{d_{\rm min}}$, as a function of (a) the dimensionless time, and (b) the number of transits for different $Ra$ values. Crosses mark the values of $\sigma_{d_{\rm min}}$ for the snapshots displayed in Figure~\ref{fig:min_depth_snapshots} for $Ra=10^{12}$. The expression of the fit common to all curves (yellow dashed curve) is given in Equation~\eqref{eq:d_min_STD}. Results of 3D simulations are shown with the dotted lines.}
\label{fig:std_min_depth_evolutions}
\end{figure}


Similarly, $\sigma_{d_{\rm min}}$, the standard deviation of the minimum depth reached by the fluid shows a decreasing exponential dependence with the number of transit times (Figure~\ref{fig:std_min_depth_evolutions}). For the different Rayleigh numbers, the standard deviation of $d_{\rm min}$ as a function of the number of transit times is well fitted by the following expression:
\begin{equation}
\sigma_{d_{\rm min}} = 0.2356 \exp[-0.3726 N_{\rm transits}] + 0.04936 \exp[-0.01616 N_{\rm transits}] .
\label{eq:d_min_STD}
\end{equation}

\noindent This fit is displayed by the yellow dashed curve in Figure~\ref{fig:std_min_depth_evolutions}b, and its expression will be used as the characteristic property of the distribution of $d_{\rm min}$ in the next Section (in Equation~\eqref{eq:Laplace_function}).


\begin{figure}
\centering
\includegraphics[width=1\textwidth, height=1\textheight, keepaspectratio]{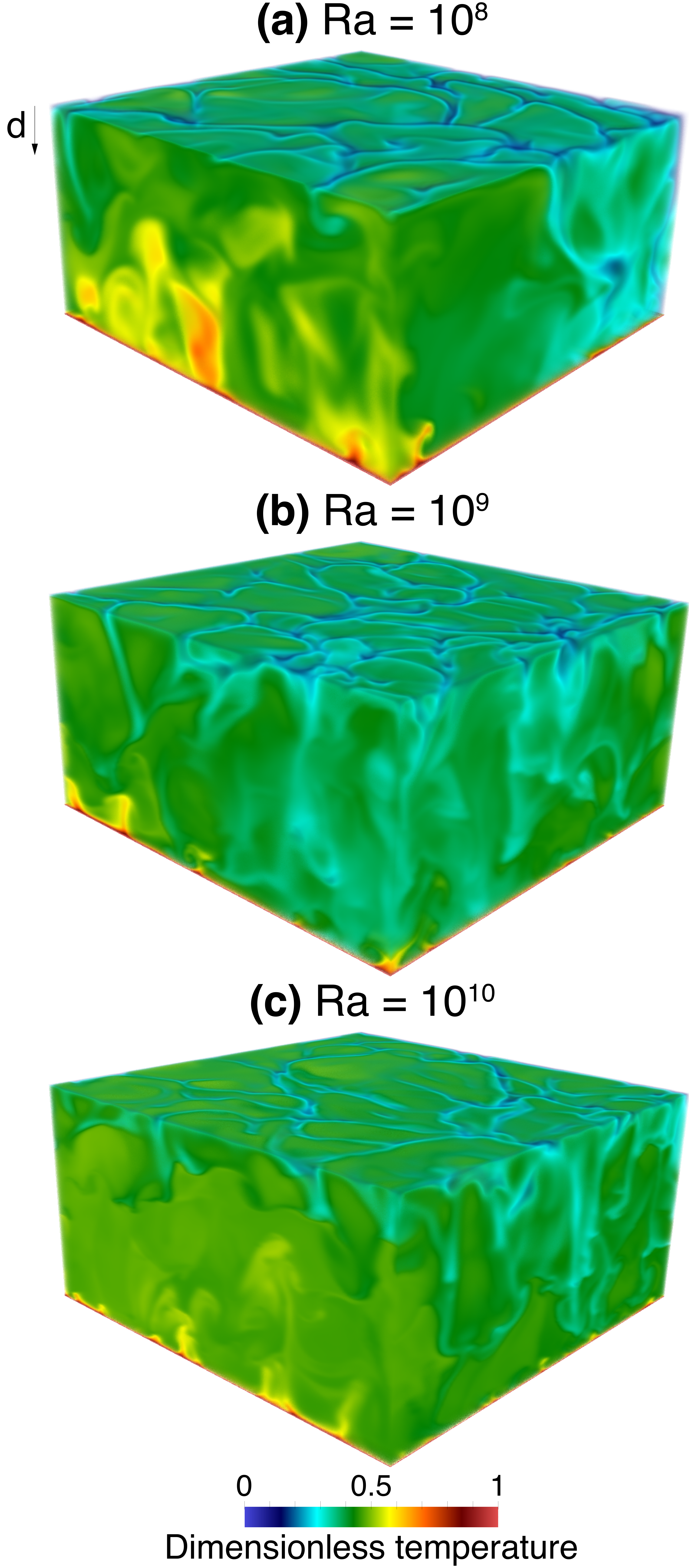}
\caption{Results of the numerical experiments in a 3D geometry. Temperature fields at statistical steady-state, at $Pr=1$, considering an aspect ratio of two in the horizontal directions.}
\label{fig:3D_temperature_field}
\end{figure}


Previous studies reported on the fact that a third spatial degree of freedom may affect both the global (\ie $Nu$ and $Re$ numbers) and local properties of the convective flow, especially at low $Pr$ number values \citep[\eg][]{Schmalzl2004, vanderPoel2013}. In particular, \cite{Schmalzl2002, Schmalzl2004} suggested that differences between 2D and 3D convection may appear as a result of (i) the constrained motion of the large-scale circulation in 2D flows compared to 3D geometry, and (ii) the increasing partitioning of energy in the toroidal component of the velocity in 3D flows at $Pr < 1$, otherwise absent in 2D geometry, thus strengthening the large-scale component of the flow \citep{Breuer2004}.
However, other parameters such as the boundary conditions, the aspect ratio, or the existence of multiple states and potential multi-stability observed around $Pr = 1$ may also play a significant, yet not sufficiently constrained, role in these discrepancies \citep[\eg][]{vanderPoel2011, vanderPoel2012, vanderPoel2013}. 
Therefore, to test the aforementioned influences on our results we performed a series of numerical experiments using a 3D geometry. These experiments considered a larger aspect ratio (of two in the horizontal directions) to ensure that the flow motions were not artificially constrained to a unit cell in our 2D experiments.
Figure~\ref{fig:3D_temperature_field} shows snapshots of the temperature field for the different Rayleigh numbers considered at statistical steady-state. The convective features and their variations with the Rayleigh number are in good agreement with the qualitative description made for the 2D simulations.
The main parameters controlling the distribution of $d_{\rm min}$ (hence the ability of the magma to reach a given exsolution depth) obtained for the 3D simulations are shown in Figures~\ref{fig:min_depth_evolutions} and~\ref{fig:std_min_depth_evolutions} (dotted lines).
As in the 2D simulations, when plotted as a function of the number of convective transits, both the spatially-averaged, $\langle{d_{\rm min}}\rangle$, and standard deviation, $\sigma_{d_{\rm min}}$, of the minimum depth reached by the tracers collapse onto a master curve for all Rayleigh numbers (Figures~\ref{fig:min_depth_evolutions}b and~\ref{fig:std_min_depth_evolutions}b), supporting the fact that the ability of the magma to reach given depths is essentially governed by the magnitude of the convective velocities, as shown in the 2D experiments. Moreover, the master curves fitting 2D data (yellow dashed lines in Figures~\ref{fig:min_depth_evolutions}b and~\ref{fig:std_min_depth_evolutions}b; Equations~\eqref{eq:d_min_MEAN} and~\eqref{eq:d_min_STD}) are in good agreement with the 3D calculations. This demonstrates that the ability of the convecting fluid to transport material to shallow depths remains comparable in 3D.
This confirms that the main quantity that affects the convecting outgassing efficiency is the magnitude of vertical velocities, regardless of the existence of toroidal motion.
The observed good agreement between 2D and 3D cases implies that we can confidently apply the scalings derived from our 2D experiments.

\subsection{Description of the global distribution of the minimum depth reached by fluid parcels}
In addition to the values of $\langle d_{\rm min} \rangle$ and $\sigma_{d_{\rm min}}$, a more general description of the distribution of the minimum depth reached by the tracers (Figures~\ref{fig:min_depth_snapshots}e to \ref{fig:min_depth_snapshots}h) is required to accurately predict the magma ocean outgassing efficiency.
As seen in Figures~\ref{fig:min_depth_snapshots}e to \ref{fig:min_depth_snapshots}h, when the number of transit times increases, the distribution of $d_{\rm min}$ becomes asymmetric as a result of convective motion. Using $\sigma_{d_{\rm min}}$ to constrain its spread, such a distribution can be approximated by a truncated asymmetric Laplace distribution, whose probability density function writes:

\begin{equation}
La(d_{\rm min}) = \frac{1}{2b}\exp{\left(-\frac{\mid d_{\rm min}-\mu\mid}{b}\right)},
\label{eq:Laplace_function}
\end{equation}

\noindent where the parameters $\mu = 0$ (leading to a truncated asymmetric distribution) and $b=\sqrt{\sigma_{d_{\rm min}}^2/2}$, with $\sigma_{d_{\rm min}}$ given by Equation~\eqref{eq:d_min_STD}.
In principle, Equation~\eqref{eq:Laplace_function} needs to be integrated between $d_{\rm min}=-\infty$ and $d_{\rm min}=+\infty$ for its integral to equal unity, \ie for encompassing the entire range of values that can be taken, which is nonphysical in our case since $d_{\rm min}$ ranges between 0 and 1. Therefore, we normalize Equation~\eqref{eq:Laplace_function} such that the probability density function equals unity when integrated for depths ranging between 0 and 1:

\begin{equation}
\mathcal{L}= \frac{La(d_{\rm min})}{\int_{d_{\rm min}=0}^{d_{\rm min}=1}La(d_{\rm min})~{\rm d}d_{\rm min}}.
\label{eq:NLaplace_function}
\end{equation}

\begin{figure}
\centering
\includegraphics[width=0.65\linewidth]{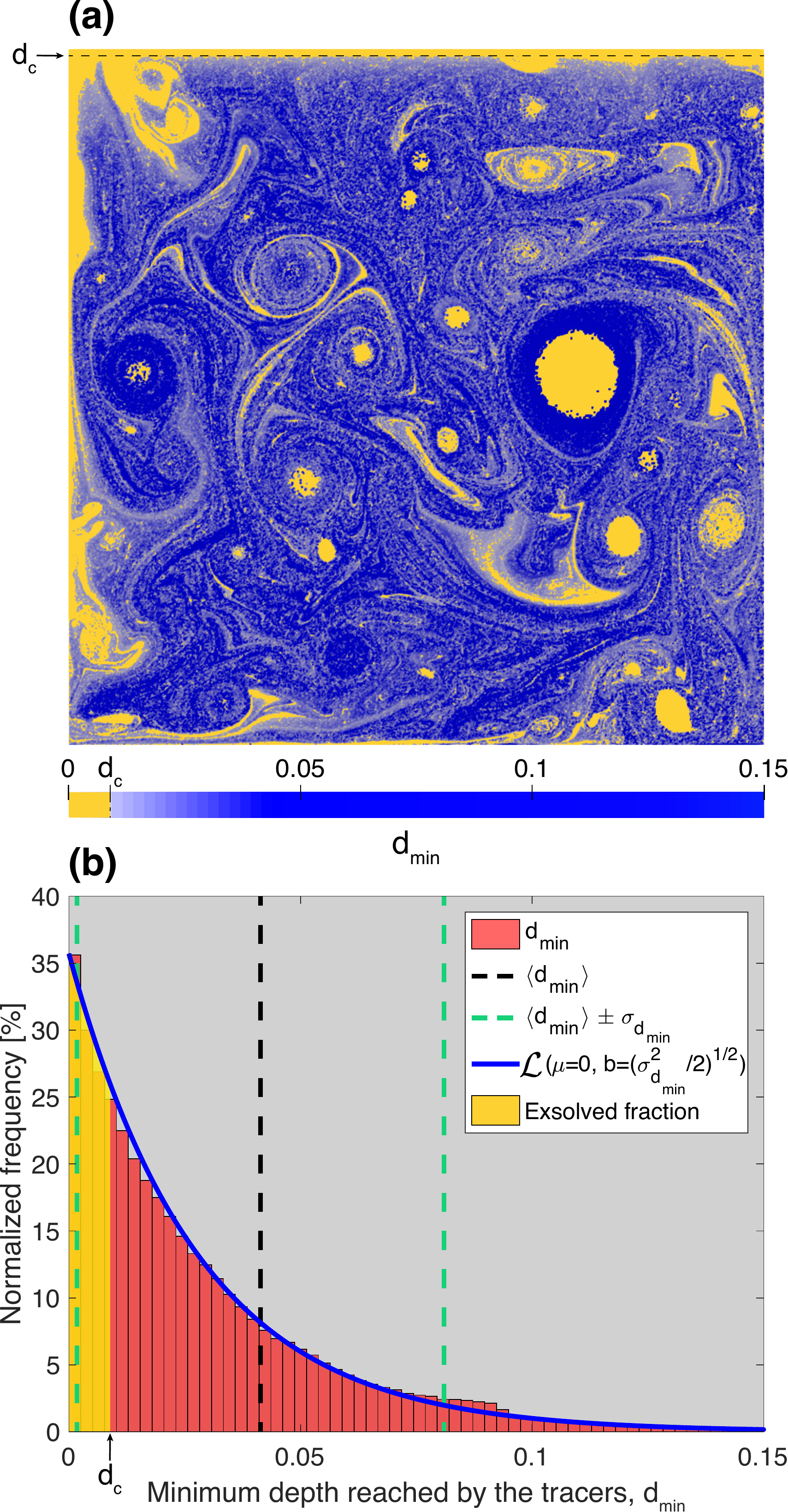}
\caption{(a) Results from the numerical experiments. Field of the dimensionless minimum depth reached by the passive Lagrangian tracers, $d_{\rm min}$, for $Ra=10^{12}$ after 15 transits, and (b) corresponding normalized distribution (red histogram). The spatially averaged $d_{\rm min}$ and standard deviation are shown by the vertical dashed black and green lines, respectively. The distribution can be approximated by the probability density function of a truncated Laplace distribution $\mathcal{L}(\mu=0,b=(\sigma^{2}_{d_{\rm min}}/2)^{1/2})$ (blue line, Equation~\eqref{eq:NLaplace_function}, using $\sigma_{d_{\rm min}}$ obtained from Equation~\eqref{eq:d_min_STD}).
(a) The yellow area represents the magma ocean exsolved fraction corresponding to the tracers that have reached a given critical depth since the initial time $t=0$, $d_c \approx 8.85\times 10^{-3}$, corresponding to 26.55 km when considering a 3000 km deep magma ocean (horizontal dashed line). (b) The corresponding theoretical value is computed through the integration of the yellow shaded area (Equation~\eqref{eq:exsolved_percentage}).}
\label{fig:min_depth_histogram}
\end{figure}

\noindent The above normalized probability density function, $\mathcal{L}$,  is shown by the blue curve in Figures~\ref{fig:min_depth_snapshots}g, \ref{fig:min_depth_snapshots}h, and \ref{fig:min_depth_histogram}b.


\subsection{Magma ocean dynamic outgassing model and simple predictions}
Equation~\eqref{eq:NLaplace_function} can  be used  to estimate the amount of fluid that has reached a certain critical depth, $d_{c}$, for a given number of transit times, by computing its integral between $d_{\rm min}=0$ and $d_{\rm min}=d_{c}$ (yellow shaded area in Figures~\ref{fig:min_depth_histogram}a and \ref{fig:min_depth_histogram}b). If we assume this critical depth to be the exsolution depth, \ie $d_{c} = d_{\rm exsol}$, we can compute the  exsolved fraction of the fluid denoted $X_{\rm exsolved}$:

\begin{equation}
    X_{\rm exsolved}=\int_{d_{\rm min}=0}^{d_{\rm min}=d_{c}}{\mathcal{L}}~{\rm d} d_{\rm min},
\label{eq:exsolved_percentage}
\end{equation}
which corresponds to the fraction of the melt that has reached the exsolution depth and thus has been able to exsolve its oversaturated volatiles.
This predicted value for $X_{\rm exsolved}$ can be compared to the   value obtained in our numerical experiments, $X^{\rm exp}_{\rm exsolved}$, given by:
\begin{equation}
    X^{\rm exp}_{\rm exsolved} = \frac{N_{\rm tracers}(d_{\rm min} \leq d_{c})}{N_{\rm tracers}},
\end{equation}

\noindent where $N_{\rm tracers}(d_{\rm min} \leq d_{c})$ is the number of tracers that have reached the critical depth.
This comparison allows one to estimate the uncertainty on the predicted magma ocean exsolved fraction computed, $\Delta X_{\rm exsolved}= X_{\rm exsolved}- X^{\rm exp}_{\rm exsolved}$, illustrated by the deviation between the theoretical Laplace distribution (blue curve in Figures~\ref{fig:min_depth_snapshots}g, \ref{fig:min_depth_snapshots}h, and \ref{fig:min_depth_histogram}b) and the observed, empirical distribution of $d_{\rm min}$ (red histogram in Figures~\ref{fig:min_depth_snapshots}g, \ref{fig:min_depth_snapshots}h, and \ref{fig:min_depth_histogram}b). Note that $\Delta X_{\rm exsolved}$ depends both on the critical depth, $d_c$, and on the number of convective transits considered. Out of 200 critical depth values for more than 50 number of transits tested, we found that the standard deviation of the error of the theoretical magma ocean exsolved fraction is $\Delta X_{\rm exsolved} = \pm 6\%$.
Even though the error was computed for $Ra=10^{12}$ only, choosing our maximum value for $Ra$ is reasonable given the fact that Rayleigh number values for global magma oceans are high.


\begin{figure}
\centering
\includegraphics[width=1.\textwidth, height=1\textheight, keepaspectratio]{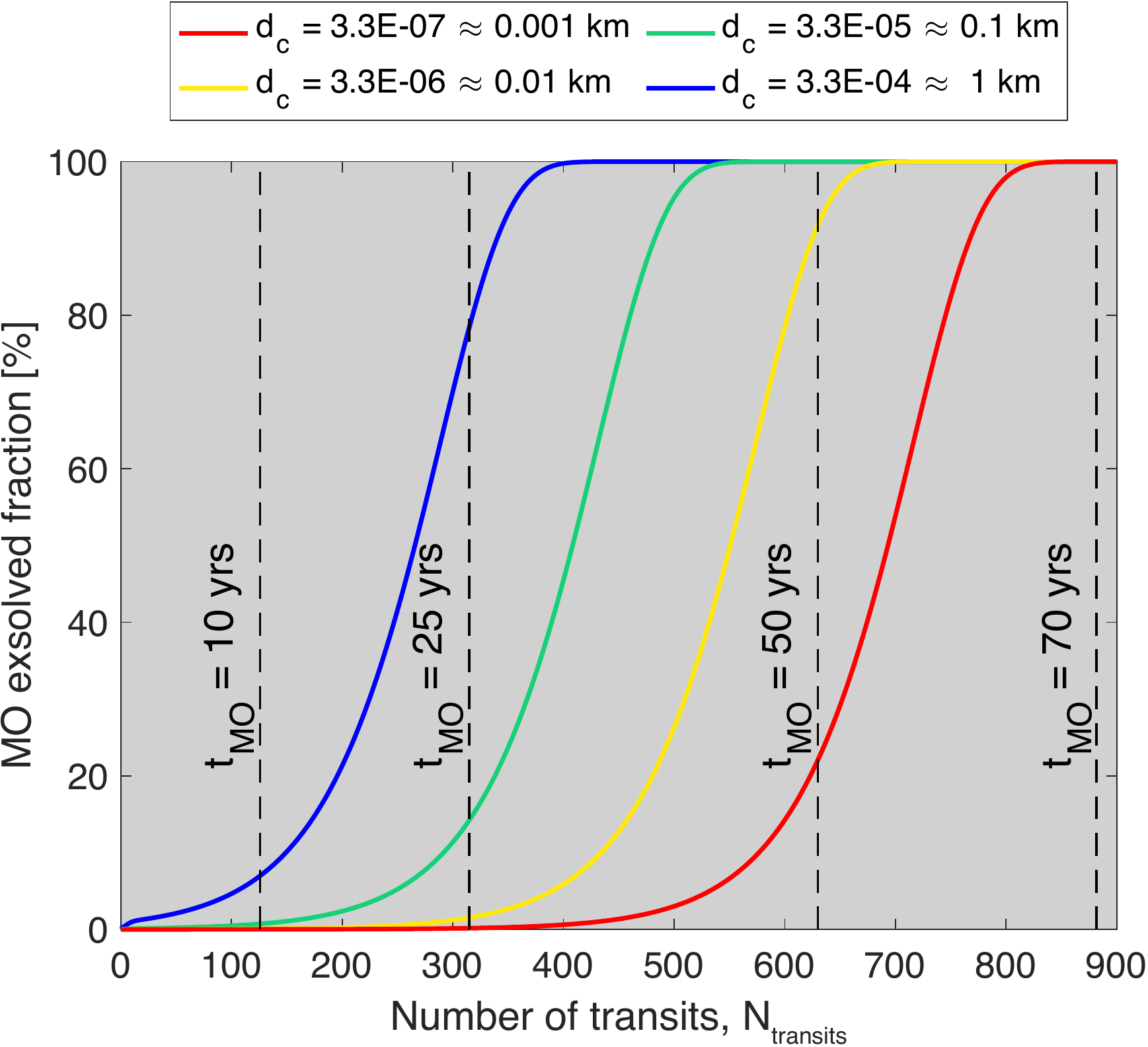}
\caption{Predicted exsolved fraction of the magma ocean as a function of the number of transits for different arbitrary critical depths $d_c$, corresponding to different exsolution depths $d_{\rm exsol}$. Duration of a vigorously convecting magma ocean transient stage corresponding to several values of number of transits are shown by the vertical dashed black lines. Standard properties for a terrestrial magma ocean are chosen (reference case) to compute these duration and are listed in Table~\ref{tab: MO_parameters}.}
\label{fig:exsolved_percentage_evolution}
\end{figure}



\begin{table}
\centering
\begin{threeparttable}
\caption{Vigorously convecting, fully molten magma ocean typical set of parameters (reference case) used to calculate the times spent by such an hypothetical Earth-sized planet magma ocean in the hard turbulent convection regime, $Ra=6.62\times 10^{29}-Pr=1$}
\label{tab: MO_parameters}
\begin{tabular}{lc}
\hline
Parameter, Symbol & Value \\
\hline
Planetary radius, $R_{\oplus}$ & $6.378\times10^6$ m \\
Melt density, $\rho$ & 4200 $\rm kg~m^{-3}$ \\
Gravity, $g$ & 9.81 $\rm m~s^{-2}$ \\
MO thickness, $L$ & $3\times10^6$ m \\
Temperature difference, $\Delta T$ & 250 K \\
Thermal capacity, $C_P$ & $10^{3}$ $\rm J~kg^{-1}~K^{-1}$ \\
Thermal conductivity, $k$ & $4$ $\rm W~m^{-1}~K^{-1}$ \\
Thermal diffusivity, $\kappa$ & $10^{-6}$ $\rm m^2~s^{-1}$ \\
Thermal expansion, $\alpha$ & $10^{-5}$ $\rm K^{-1}$ \\
Angular velocity, $\Omega$ & $10^{-4}$ $\rm s^{-1}$ \\
Dynamic viscosity, $\eta$ & $\rho\times \kappa$ \\
\hline
\end{tabular}
\end{threeparttable}
\end{table}


Figure~\ref{fig:exsolved_percentage_evolution} displays the predicted exsolved fraction of a magma  ocean, $X_{\rm exsolved}$, as a function of the number of transits achieved, for different critical depths, $d_c$, that can be considered as arbitrary exsolution depths, $d_{\rm exsol}$. Note that the exsolution depth of volatile species inversely varies with its solubility in silicate melts so that larger exsolution depths would correspond to less soluble volatile species. The values of the parameters used to compute the different curves correspond to those of the reference case listed in Table~\ref{tab: MO_parameters}. Figure~\ref{fig:exsolved_percentage_evolution} shows that $X_{\rm exsolved}$ strongly depends on the critical depth considered. Indeed, the number of transits required to exsolve 100\% of the magma ocean ranges from about 425 for $d_c = 1~\rm{km}$, to about 850 transits for $d_c=0.001~\rm{km}$ (blue and red curves, respectively). This demonstrates that a substantial number of convective transits can be required for the oversaturated volatiles to be fully exsolved. Consequently, for several values of $d_c$, the exsolution of oversaturated volatiles and therefore their degassing may not be considered to be essentially instantaneous, contrary to what is generally assumed during each computational time step in magma ocean modeling studies \citep{Elkins-Tanton2008,Hamano2013,Lebrun2013,Hamano2015,Schaefer2016,Massol2016,Salvador2017,Nikolaou.etal2019,Bower2019,Lichtenberg.etal2021,Barth2021}.
The times spent in the magma ocean transient stage associated to parameters given in Table~\ref{tab: MO_parameters} are shown for comparison (vertical dashed lines in Figure~\ref{fig:exsolved_percentage_evolution}). Details on the relationship between the number of transits and the duration of a given magma ocean transient stage, $t_{\rm MO}$, are given in the next Section.
For the hypothetical 3000 kilometers deep magma ocean considered in our reference case (Table~\ref{tab: MO_parameters}), about 35 years are sufficient to fully outgas the magma ocean if the exsolution depth is $d_{\rm exsol}\approx 1$ km, while almost 69 years are required if $d_{\rm exsol}\approx 0.001$ km (Figure~\ref{fig:exsolved_percentage_evolution}, blue and red curves, respectively). Indeed, large exsolution depths require less transit times to be reached by the entire fluid.
This implies that the degassing efficiency of magma oceans increases with the exsolution depth, and with the duration of the magma ocean stage (\ie the entire melt volume has more time to reach the exsolution depth).
Importantly, these timescales are much larger than what coupled magma ocean--atmosphere models typically assume for complete circulation of the melt up to the surface (based on high convective velocities only), and hence exsolution and degassing of all oversaturated volatiles, to proceed \citep[in about one to three weeks; \eg][]{Elkins-Tanton2008}.
As the magma ocean experiences successive different transient stages during its solidification, from a vigorously convecting fully molten mantle to a solid-state slowly convecting mantle, each of them being associated with significantly different parameters (such as paired values of $Ra$ and $Pr$ numbers, convective velocities, magma ocean depths) and mixing rates, the exsolved fraction speed rates will considerably vary during the entire evolution sequence \citep{Lebrun2013,Salvador2017,Nikolaou.etal2019}.
For the time spent in a given transient stage (corresponding to a given set of $Ra$ and $Pr$ numbers), magma ocean degassing efficiency can be estimated by comparing the time required to fully exsolve the magma ocean in such transient stage, which we define as the time required to reach $X_{\rm exsolved}=100\%$, with the time spent in the transient stage considered, $t_{\rm MO}$. This indicates that constraining the outgassing of a magma ocean through its entire lifetime requires the evaluation of the relevant outgassing efficiency for each transient stage encountered during the entire magma ocean evolution sequence (\ie for each set of $Ra$ and $Pr$ numbers experienced during the entire cooling sequence). 
While the above estimates of mama ocean degassing efficiency remain crude and based on arbitrary exsolution depths, more accurate outgassing efficiency predictions based on realistic exsolution depths for water and magma ocean dynamics are discussed in the following Section, focusing on highly vigorous stages.

\section{Application to Outgassing in Vigorously Convecting Silicate Planetary Magma Oceans \label{sec: discussion}}

In the previous Section our estimates of the magma ocean outgassing efficiency rely on two sets of parameters arbitrarily chosen: (1) the parameters that govern the magma ocean dynamics considered to estimate the time spent by the magma ocean in a given regime, $t_{\rm MO}$ (chosen here to correspond the hard turbulent convection regime, see Table~\ref{tab: MO_parameters}), and (2) the volatile species exsolution depth, $d_{\rm exsol}$.
Note that we do not consider the \textquotedblleft{}ultimate\textquotedblright{} convective regime proposed in \eg \citet{Kraichnan1962, Chavanne1997, Lepot2018} because the existence of this regime remains controversial \citep[\eg][]{Lohse2003, Roche2010}.
Depending on the values of these sets of parameters, the degree of magma ocean outgassing changes significantly, because the time required to exsolve all the oversaturated volatile species at a given evolution stage may vary over several orders of magnitude.
These two parameters are self-consistently computed in several magma ocean thermal evolution models \citep[\eg][]{Elkins-Tanton2008,Hamano2013,Lebrun2013,Hamano2015,Massol2016,Schaefer2016,Salvador2017, Nikolaou.etal2019, Bower2019,Lichtenberg.etal2021,Barth2021}. While such models could  be used together with the relationships derived in this current study to compute a more accurate magma ocean outgassing sequence by considering a realistic exsolved fraction along each transient stage of the thermal evolution, this is out of the scope of this paper. Nevertheless, in what follows, we consider the effect of realistic exsolution depths and magma ocean dynamics along with their associated time scales to make reasonable predictions of volatile outgassing in vigorously convecting planetary magma oceans. 

\subsection{Estimation of the exsolution depths}
Realistic exsolution depths can be determined out of solubility laws, as a function of the abundance of volatile species present in the melt. These laws give the lithostatic saturation pressure of a given volatile species as a function of its dissolved amount within the melt, which is usually expressed in terms of volatile mass fraction relative to the mass of the melt, $X^{\rm MO}_{\rm vol}$.
Thus, they control the partitioning of a volatile species between the silicate melt and gas phase according to a power-law relationship between saturation pressure and volatile abundance in the magma following \citep[\eg][]{Carroll1994, Holloway1994, Papale1997, Wallace1999}:
\begin{equation}
P^{\rm sat}_{\rm vol}(X_{\rm vol}^{\rm MO})=\left( \frac{X_{\rm vol}^{\rm MO}}{\zeta_{\rm vol}} \right)^{\beta_{\rm vol}},
\label{eq:solub_law}
\end{equation}

\noindent where $X_{\rm vol}^{\rm MO} = M_{\rm vol}/M_{\rm MO}$ is the mass fraction of the volatile species expressed as the mass of the volatile species within the melt, $M_{\rm vol}$, relative to the mass of the melt, $M_{\rm MO}$, $\zeta_{\rm vol}$ is a constant, and $\beta_{\rm vol}$ is the power law coefficient (see Table~\ref{tab: solub_law} for values considered here).
For a given species, the saturation pressure, $P^{\rm sat}_{\rm vol}$, is the overburden pressure threshold below which (above in terms of depth) no more volatile can be dissolved within the melt, \ie the maximum amount of dissolved volatiles within the liquid is reached, and any extra oversaturated volatile is then extracted out of the solution (\ie \textquotedblleft{}exsolved\textquotedblright{}) to form \textit{gas} bubbles.
In coupled magma ocean--atmosphere studies, the saturation pressures computed out of solubility laws are directly considered to be equivalent to the atmospheric gas partial pressures, \ie $P^{\rm sat}_{\rm vol} = P^{\rm atm}_{\rm vol}$, and thus sometimes referred to as the gas pressures within the melt, and are used to compute the corresponding atmospheric mass of the volatile species, $M_{\rm vol}^{\rm atm}$, following \citep[\eg][]{Elkins-Tanton2008,Pierrehumbert2010_book,Hamano2013,Lebrun2013,Hamano2015,Schaefer2016,Salvador2017, Nikolaou.etal2019, Bower2019,Lichtenberg.etal2021,Barth2021}:
\begin{equation}
    M_{\rm vol}^{\rm atm} = \frac{4\pi R^{2}_{p}}{g} \left(\frac{\mu_{\rm vol}}{\bar{\mu}}\right) P^{\rm sat}_{\rm vol}(X_{\rm vol}^{\rm MO}),
\end{equation}
where $\mu_{\rm vol}/\bar{\mu}$ is the ratio of the molar mass of the volatile species to the mean molar mass of the atmosphere, which equals one when considering single-component atmospheres and is thus omitted in such a case. As discussed earlier, this direct correspondence between the volatile species mass fraction in the magma ocean and its partial pressure in the atmosphere, and the resulting implicit partial pressures equilibrium between the gas phase in the melt and the atmosphere implies the instantaneous, and thus efficient, outgassing of any oversaturated volatile species at each computational time step (efficient outgassing hypothesis), thus neglecting the effect of magma ocean convective dynamics on the amount of melt actually reaching the exsolution depth and able to form gas bubbles. Again, here we revisit this assumption of efficient and instantaneous outgassing by realistically computing the amount of melt reaching the exsolution depth as a function of time.
The depth at which volatile species are exsolved out of the melt and start forming gas bubbles is thus the exsolution depth, $d_{\rm exsol}$, which is calculated from the corresponding saturation pressure. Assuming hydrostatic equilibrium, the saturation pressure of a given volatile species, $P^{\rm sat}_{\rm vol}$, is:

\begin{equation}
    P^{\rm sat}_{\rm vol}(X_{\rm vol}^{\rm MO}) = \rho ~g ~ d_{\rm exsol}.
    \label{eq:hydrostat_eq}
\end{equation}


\begin{table}
\centering
\begin{threeparttable}
\caption{Parameters determining the partitioning of water between the melt and the gas phase to be used in Equation~\eqref{eq:solub_law}}
\label{tab: solub_law}
\begin{tabular}{l c c}
\hline
Parameter & Value & Reference \\
\hline
Exponent $\beta_{\rm H_2O}$  & 1/0.7                                   & \citet{Carroll1994} \\
Constant $\zeta_{\rm H_2O}$     & $6.8\times10^{-2}\, {\rm ppm~Pa}^{-1}$  & \citet{Carroll1994} \\
\hline
\end{tabular}
\end{threeparttable}
\end{table}


For the sake of simplicity, the above quantity is assumed to be constant for a given magma ocean transient stage (and corresponding pair of $Ra-Pr$ numbers) at which outgassing efficiency is evaluated.
Two main mechanisms favor volatile exsolution during magma ocean cooling. For a given transient stage, in a convective mantle upwelling, the ambient overburden (lithostatic) pressure of an ascending volume of melt decreases, and as soon as the saturation pressure is reached, volatiles in excess of saturation exsolve out of the melt and form gas bubbles (Figures~\ref{fig:degassing_sketch} and \ref{fig:solubility_laws}) \citep[\eg][]{Lesher2015}.
Furthermore, volatiles are incompatible species: during melt crystallization they partition preferentially in the liquid rather than in the solid phase. Thus, in evolving magmatic systems such as magma oceans, volatiles concentrate into the melts, which become more and more enriched in volatiles, such that saturation occurs more easily and at increasing depths with cooling and crystallization leading to enhanced volatile species exsolution with time (Figure~\ref{fig:solubility_laws}).

The coexistence of several volatile species dissolved into the melt, such as \ce{H2O} and \ce{CO2}, influence their respective solubility, leading to more complex relationships between the volatile abundances and saturation pressures \citep[\eg][]{Papale1999, Berlo2011, Massol2016}.
For the sake of simplicity, we will restrict our analysis to the case of a single-component atmosphere, made of pure water vapor \ce{H2O}, which is one of the most important greenhouse gas and plays a fundamental role in planetary habitability \citep[\eg][]{Wallace2006_book, Pierrehumbert2010_book,Westall2018}.
For the solubility of water within the magma ocean, we use the law classically referred to as \citet{Carroll1994} (whose parameters are given in Table~\ref{tab: solub_law}), which is commonly used in coupled magma ocean--atmosphere studies \citep[\eg][]{Lebrun2013, Salvador2017, Nikolaou.etal2019, Bower2019}. Implications of other water solubility laws are discussed in \ref{solubility_laws}. Although the uncertainty related to the choice of the solubility law is not negligible, it would not affect the conclusions.
We assume a given water budget, expressed as a number of current Earth ocean mass that translates to a given mass fraction of dissolved water relative to the melt mass, $X_{\rm \ce{H2O}}^{\rm MO}$, which, together with the solubility law (Equation~\eqref{eq:solub_law}), constrains the value of the exsolution depth, $d_{\rm exsol}$ (Equation~\eqref{eq:hydrostat_eq}).

\subsection{Magma ocean governing parameters and convective timescales}
The number of transits achieved, $N_{\rm transits}$, relates to the time spent in a given magma ocean transient stage, $t_{\rm MO}$, following:
\begin{equation}
    t_{\rm MO} = N_{\rm transits} \times \tau_{\rm transit},
    \label{eq:t_MO}
\end{equation}

\noindent where $\tau_{\rm transit}$ is the transit timescale corresponding to the time required for a parcel of the convecting fluid to travel from one horizontal surface to the other, which is written:

\begin{equation}
    \tau_{\rm transit} = \frac{L}{U_{\Omega}},
\end{equation}

\noindent where $L$ is the thickness of the molten layer, and $U_{\Omega}$ refers to the convective velocity accounting for the effect of planetary rotation. The latter may influence convection and the magnitude of the associated fluid motion \citep{Solomatov2000,Maas2015,Maas2019}. Therefore, following \citet[][Equation (11)]{Solomatov2000}, this effective convective velocity is written:

\begin{equation}
    U_{\Omega} = v_{\rm conv} \left( \frac{\alpha g F}{\rho C_P \Omega} \right)^{1/2} ,
    \label{eq:U_rotation}
\end{equation}

\noindent where $\Omega$ is the planet's angular velocity, $C_P$ is the specific heat at constant pressure, $v_{\rm conv} = Re \left( \rho L /\eta \right)^{-1}$ is the convective flow velocity computed using our $Re-Ra$ scaling (Equation~\eqref{eq:Re_scaling}), and $F$ is the convective heat flux at the surface of the magma ocean:

\begin{equation}
F = Nu ~ \frac{k\Delta T}{L},
\end{equation}

\noindent where $Nu$ is the Nusselt number  computed using Equation~\eqref{eq:Nu_scaling} and $k$ is the thermal conductivity. The values of the different parameters introduced above are listed in Tables~\ref{tab: MO_parameters} and \ref{tab: MO_parameters_SE}.
\cite{Solomatov2000} estimated an uncertainty of a factor 3 on the hard turbulence convective velocities accounting for the effect of rotation, \ie $[ U_{\Omega}/3;~ 3 ~U_{\Omega}]$. 
In addition to the plausible range of convective velocities, we considered the uncertainty on two poorly constrained parameters that affect the magma ocean dynamics: its thermal expansion, $\alpha$, and the temperature difference across the upper thermal boundary layer, $\Delta T$. To account for the possible range of variation of these three parameters ($U_{\Omega}$, $\alpha$, and $\Delta T$) and their influence on the magma ocean dynamics, we considered two end-member cases in addition to our reference case: (i) an \textquotedblleft{}enhanced degassing limit\textquotedblright{}, where convection is the most efficient due to high convective velocities ($3 ~ U_{\Omega}$), a large temperature difference across the thermal boundary layer ($\Delta T = 1000$ K; \eg \citealp{Lebrun2013}), and a large thermal expansion ($\alpha = 10^{-5}~\rm K^{-1}$), and (ii) a \textquotedblleft{}weak degassing limit\textquotedblright{} associated with less vigorous convection, and with smaller convective velocities ($ U_{\Omega}/3$), smaller temperature difference ($\Delta T = 20$ K; \eg \citealp{Lebrun2013}), and smaller ($\alpha = 10^{-6}~\rm K^{-1}$) thermal expansion (Table~\ref{tab: degassing_cases_parameters}).


\begin{table}
\centering
\begin{threeparttable}
\caption{Vigorously convecting, fully molten magma ocean set of parameters considered for the Super-Earth reference case and differing from those given in Table \ref{tab: MO_parameters}, $Ra=5.52\times 10^{30}-Pr=1$}
\label{tab: MO_parameters_SE}
\begin{tabular}{lc}
\hline
Parameter, Symbol & Value \\
\hline
Planet to Earth mass ratio, $M_p/M_{\oplus}$ & 5 \\
Planetary radius\textsuperscript{*}, $R_p$ & $10.65\times10^6$ m \\
Gravity, $g$ & 17.6 $\rm m~s^{-2}$ \\
MO thickness, $L$ & $L_{\oplus}R_p/R_{\oplus}$ \\
\hline
\end{tabular}
\textsuperscript{*}Using the mass-radius relationship given in \cite{Kopparapu2014} and based on the \href{http://www.exoplanets.org}{exoplanets.org} known exoplanets database \citep{Wright2011}.
\end{threeparttable}
\end{table}



\begin{table}
\centering
\begin{threeparttable}
\caption{Parameters used for the enhanced and weak degassing limit cases for both Earth-sized and super-Earth planets}
\label{tab: degassing_cases_parameters}
\begin{tabular}{lcc}
\hline
Parameter, Symbol & Enhanced degassing limit & Weak degassing limit \\
\hline
Temperature difference, $\Delta T$ & 1000 K & 20 K \\
Thermal expansion, $\alpha$ & $10^{-5}$ $\rm K^{-1}$ & $10^{-6}$ $\rm K^{-1}$ \\
Convective velocity prefactor & 3 & 1/3 \\
\hline
\end{tabular}
\end{threeparttable}
\end{table}


We compared the time required to fully exsolve all the oversaturated water from the magma ocean for the three different cases to the time spent by the magma ocean in a given transient stage, $t_{\rm MO}$. To do so, we evaluate the degassing efficiency of the global, fully molten magma ocean (corresponding to our transient stage) and consider the pair of $Ra$ and $Pr$ numbers and associated parameters as constant, for a given water content. This corresponds to the earliest and most vigorously convecting magma ocean stage described in Figure~\ref{fig:MO_stages}b. Coupled magma ocean--atmosphere models \citep[\eg][]{Solomatov2000,Lebrun2013,Nikolaou.etal2019} show that depending on the ability of the growing atmosphere to efficiently loose heat to space, the duration of such a deep, vigorously convecting, fully molten stage can vary from $t_{\rm MO}{\rm min}\approx 10$ years, when no atmosphere is present, to up to $t_{\rm MO}{\rm max}\approx 500$ years when a thick and thermally opaque atmosphere overlies the magma ocean and buffers the heat loss to space \citep[\eg][]{Lebrun2013}. Considering these two end-member scenarios, the complete solidification of the terrestrial magma ocean, commonly defined as the time at which the rheological transition front reaches the surface (\ie when the \textquotedblleft{}mush\textquotedblright{} stage is reached; panels a and d in Figure~\ref{fig:MO_stages}) ranges from a thousand years to several millions years \citep[\eg][]{Elkins-Tanton2008,Lebrun2013, Hamano2013, Hamano2015, Salvador2017, Nikolaou.etal2019, Bower2019,Krissansen-Totton2021}. Note that if one extends the atmospheric composition to other species and/or abundances, or consider different stellar, atmospheric, surface, and mantle conditions, this time interval could be even larger, and complete solidification could require more than a hundred million years \citep[\eg][]{Zahnle1988, Abe1993c, Lupu2014, Hamano2015, Schaefer2016, Monteux2016, Bonati2019, Zahnle2020, Lichtenberg.etal2021, Barth2021}.

\subsection{Results: outgassing model predictions}
The outgassing efficiency of a magma ocean depends on the duration of the magma ocean stage in the considered transient stage, and on the ability of convective motions to bring molten material to shallow depths at which the oversaturated volatiles can exsolve within such a time interval. We considered these two aspects as described below.
In addition, we make the assumption that water rapidly reaches the surface once exsolved, without being recycled/ingassed back deeper in the mantle by downwelling currents, and we neglect the influence of  water concentration on magma ocean viscosity.


\begin{figure}
\centering
\includegraphics[width=1.\textwidth, height=0.8\textheight, keepaspectratio]{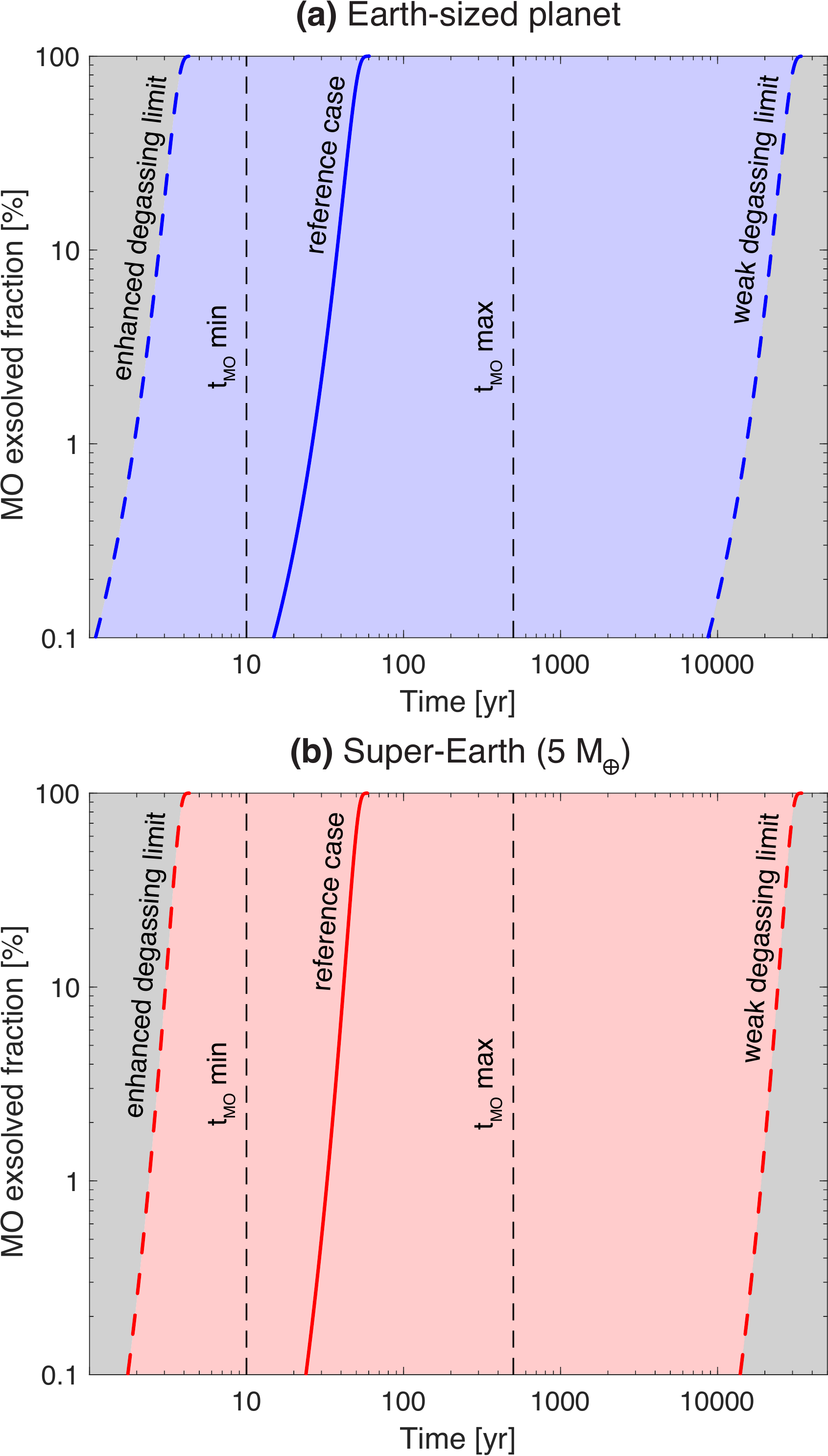}
\caption{Theoretical model predictions. Plausible range of the magma ocean fraction that has reached the exsolution depth as a function of time for (a) an Earth-sized planet (blue shaded area), and (b) a 5 Earth mass super-Earth (red shaded area). The exsolution depth is computed considering the mass fraction of 1 Earth ocean of water dissolved within a magma ocean of thickness an Earth-like planet-to-mantle radius ratio. Reference cases (plain lines) parameters for the Earth-sized planet and super-Earth are given in Tables \ref{tab: MO_parameters} and \ref{tab: MO_parameters_SE}, respectively. Parameters considered for the enhanced and weak degassing limits (blue and red dashed lines) are listed in Table~\ref{tab: degassing_cases_parameters}. Vertical black dashed lines indicate the minimum and maximum duration of the magma ocean in the fully molten transient stage from \citet{Lebrun2013} coupled magma ocean--atmosphere model.}
\label{fig:Xexsolved_compa_TP_VS_SE_1Meo}
\end{figure}


\subsubsection{Influence of the convective strength and of the planet size}
Figure~\ref{fig:Xexsolved_compa_TP_VS_SE_1Meo} shows the magma ocean exsolved fraction as a function of the time spent in the fully molten state for (a) an Earth-sized planet and (b) for a 5 Earth-mass planet. The corresponding magma ocean parameters are listed in Tables~\ref{tab: MO_parameters} and \ref{tab: MO_parameters_SE}, respectively, for the reference cases (plain blue and red curves, respectively), and in Table~\ref{tab: degassing_cases_parameters} for the enhanced and for the weak degassing limits (dashed blue and red curves, respectively). The magma ocean thickness for the super-Earth case is scaled based on Earth's mantle-to-planet radius ratio.
For both planets, the water content is fixed to one Earth ocean mass, $1~M_{\rm EO}=1.4\times 10^{21}~\rm kg$, initially dissolved and homogeneously distributed within the magma ocean.
For the 3000 km deep Earth-sized magma ocean, this corresponds to a mass fraction of dissolved water relative to the magma ocean mass equal to $X^{\rm MO}_{\rm H_2O} = 100\times \frac{1~M_{\rm EO}}{M_{\rm MO}} = 3.6\times10^{-2}~\rm wt\%$, which yields an exsolution depth, $d_{\rm exsol}$, of $5\times10^{-3}$ km (Equations~\eqref{eq:solub_law} and \eqref{eq:hydrostat_eq}).
For the 5 Earth mass rocky planet, $X^{\rm MO}_{\rm H_2O} = 7.73\times10^{-3}~\rm wt\%$, \ie the same mass of water  diluted within a larger magma ocean volume corresponds to a smaller exsolution depth $d_{\rm exsol} = 3.14\times10^{-4}~\rm km$.
The plausible volatiles exsolved fraction range corresponding to the plausible variation range of parameters explored ($U_{\Omega}$, $\alpha$, and $\Delta T$) lies within the blue and red areas of Figures~\ref{fig:Xexsolved_compa_TP_VS_SE_1Meo}a and \ref{fig:Xexsolved_compa_TP_VS_SE_1Meo}b, respectively.

For the reference cases (Tables~\ref{tab: MO_parameters} and \ref{tab: MO_parameters_SE}), 767 and 969 convective transits are required for the entire magma oceans volumes to reach the corresponding exsolution depths (Equation~\eqref{eq:exsolved_percentage}), \ie for the entire oversaturated water to be exsolved and outgassed, for the Earth-sized planet and for the super-Earth planet, respectively.
This corresponds to 61 and 59 years (Equation~\eqref{eq:t_MO}), respectively.
Despite significantly different exsolution depths, the resulting time required to fully exsolve oversaturated \ce{H2O} from the melt is similar. This is due to the fact that the larger magma ocean thickness and gravity on larger planets increase the $Ra$ number, leading to more vigorous convective motions on super-Earths. This offsets the effect of shallower exsolution depths (requiring more convective transits to be achieved to outgas volatiles), and results in a more efficient degassing on super-Earths. Overall, magma ocean exsolution starts slightly later on super-Earths than on Earth-sized planets, but full degassing is reached after similar amounts of time spent in the considered magma ocean transient stage, $t_{\rm MO}$, leading to a slightly faster magma ocean degassing on super-Earths.

For both types of planets (Figures~\ref{fig:Xexsolved_compa_TP_VS_SE_1Meo}a and \ref{fig:Xexsolved_compa_TP_VS_SE_1Meo}b, respectively), the timing of degassing for the reference case lies between the shortest and longest magma ocean fully molten stage duration [$t_{\rm MO} {\rm min}$; $t_{\rm MO} {\rm max}$]. This implies that none of the potentially available oversaturated water would have been exsolved and outgassed in the fast cooling case ($t_{\rm MO} {\rm min}$), while it would be fully exsolved and outgassed in the slow cooling scenario ($t_{\rm MO} {\rm max}$). Furthermore, this indicates that for a conservative set of magma ocean governing parameters, complete or partial outgassing remain possible. 
However, if one considers a weak degassing limit, exsolution never starts before the end of the magma ocean transient stage considered, even for its longest duration, $t_{\rm MO} {\rm max}$, regardless of the planet size (Figures~\ref{fig:Xexsolved_compa_TP_VS_SE_1Meo}a-b). In contrast, the enhanced degassing limit shows the opposite behavior since the complete outgassing always occurs before the end of the magma ocean transient stage considered, even for its shortest duration, $t_{\rm MO} {\rm min}$, associated with the fastest cooling rate (Figures~\ref{fig:Xexsolved_compa_TP_VS_SE_1Meo}a-b).
Yet, the cooling rate itself is controlled by the amount of greenhouse gases in the atmosphere and thus by the outgassing rate. Therefore, if the outgassing is not efficient and slow as in the weak degassing limit scenario or for small planets, the cooling rate will be high, allowing less time for the magma ocean to outgas, thus making the degassing efficiency weaker and so on. Conversely, for the fast, enhanced degassing limit scenario and for large planets, the amount of greenhouse gases outgassed will slow down the solidification, making the degassing more efficient by increasing the duration of the magma ocean transient stages and leaving more time for the melt to reach the exsolution depth.

Therefore, between the extreme enhanced and weak degassing limits, various combinations of parameters governing magma ocean dynamics  would result in either a fast and complete outgassing, an incomplete degassing, or no degassing at all.
These results emphasize the fact that magma ocean degassing strongly depends on its dynamics and associated parameters, and underline once again that complete outgassing during the most vigorously convecting, fully molten magma ocean stage (and by extension, during subsequent, less vigorous transient stages) is far from being certain even for magma oceans lasting for several thousands of years in the fully molten state (Figures~\ref{fig:Xexsolved_compa_TP_VS_SE_1Meo}a-b). 

Since we assumed that the same amount of water is dissolved within a smaller magma ocean volume, the Earth-sized magma ocean is more concentrated in water compared to that of the super-Earth case, as it would be in a super-Earth magma ocean at an advanced crystallization stage (because water partitions preferentially into the remaining melt, whose volatile concentration increases with solidification). This, in addition to its smaller magma ocean thickness, Rayleigh number, and associated convective vigor, makes the Earth-sized case the analog of a more advanced crystallization stage than the super-Earth case.
While we focused on the most vigorous fully molten stage, similar outgassing timings between these two different planetary cases, that would be representative of two slightly different convection efficiency/crystallization stages, suggest that our conclusions would apply for later transient crystallization stages and associated less vigorous convective dynamics (characterized by other pairs of $Ra-Pr$ values experienced during subsequent cooling stages). This suggests that degassing efficiency would decrease with time, along the solidification of the magma ocean due to the decrease of its convective vigor buffering the increase of the exsolution depth associated with water enrichment of the melt.


\begin{figure}
\centering
\includegraphics[width=\textwidth, height=1.2\textheight, keepaspectratio]{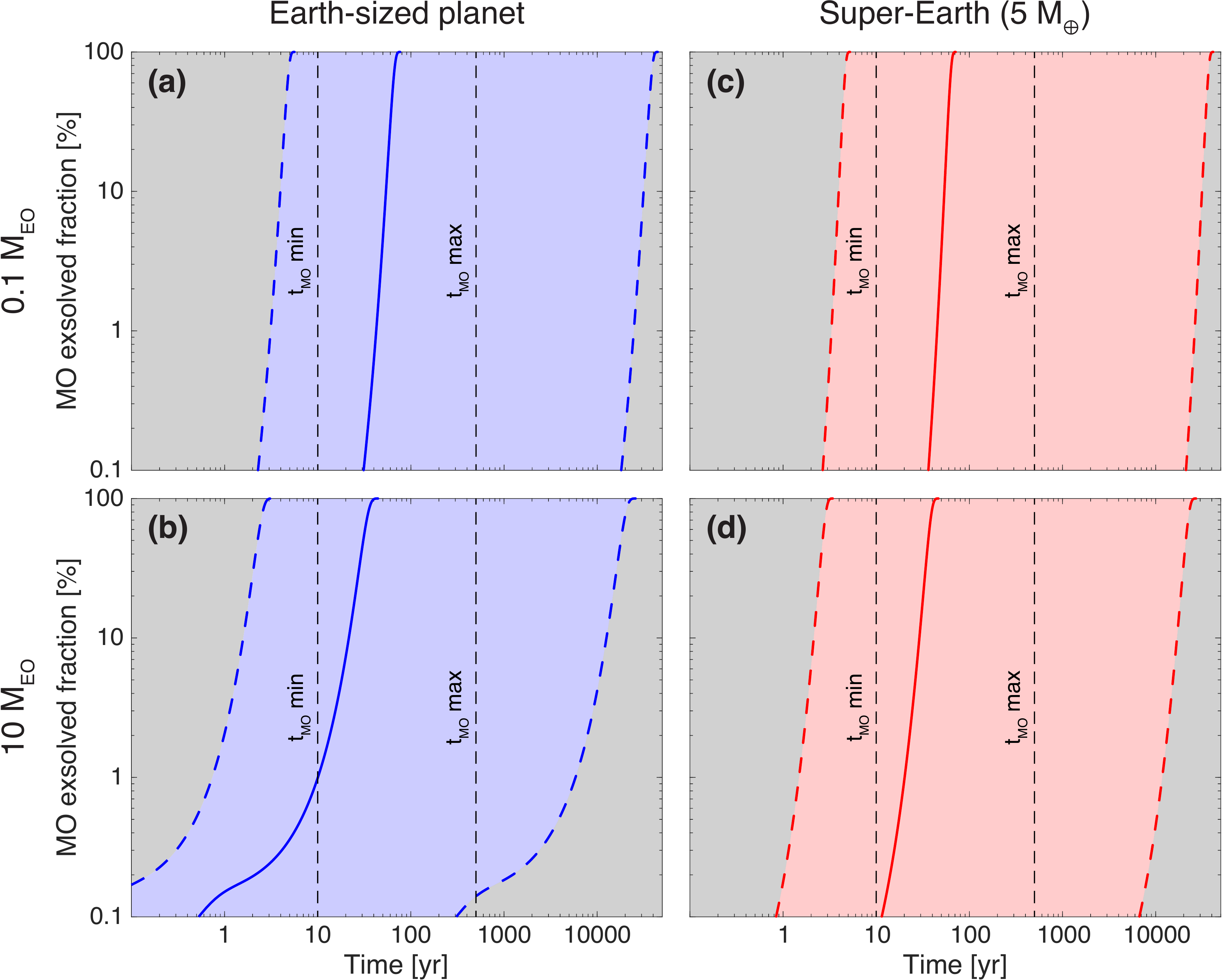}
\caption{Theoretical model predictions. Plausible range of the magma ocean exsolved fraction as a function of time on (a-b) an Earth-sized planet (blue shaded area), and (c-d) a super-Earth (red shaded area), for an exsolution depth out of (a, c) 0.1 Earth ocean, and (b, d) 10 Earth oceans dissolved within the magma ocean. The values of the governing parameters for the reference cases (plain lines) for the Earth-sized planet and super-Earth are listed in Tables~\ref{tab: MO_parameters} and \ref{tab: MO_parameters_SE}, respectively. Parameters considered for the enhanced and weak degassing limits (blue and red dashed lines) are listed in Table~\ref{tab: degassing_cases_parameters}. Vertical black dashed lines indicate the minimum and maximum duration of the magma ocean in the fully molten transient stage from \citet{Lebrun2013} coupled magma ocean--atmosphere model.}
\label{fig:Xexsolved_compa_TP_SE_01Meo_VS_10Meo}
\end{figure}


\subsubsection{Influence of the initial volatile content}
Figure~\ref{fig:Xexsolved_compa_TP_SE_01Meo_VS_10Meo} compares the exsolved fractions of oversaturated water as a function of time for two different initial water contents: 0.1 and 10 Earth ocean mass, for the Earth-sized and the super-Earth cases.
For both types of planets, the higher the concentration of dissolved water, the earlier degassing starts. Indeed, larger concentrations of dissolved water result in larger saturation pressures and exsolution depths (Equation~\eqref{eq:solub_law} and Figure~\ref{fig:solubility_laws}), while other parameters governing the dynamics are unaffected (under the assumption that the liquid-state mantle viscosity is not  influenced by the dissolved water content). It is therefore easier and faster for the entire magma ocean volume to travel up and to reach these deeper exsolution levels, while shallow exsolution depths closer to the surface require more transit times to be reached (Figures~\ref{fig:Xexsolved_compa_TP_SE_01Meo_VS_10Meo}b, \ref{fig:Xexsolved_compa_TP_SE_01Meo_VS_10Meo}d, and \ref{fig:exsol_time_VS_exsol_depth}). For completeness, the influence of the exsolution depth on the exsolution time is discussed in \ref{dexsol_influence}.
This behavior is even more pronounced on highly water-concentrated Earth-sized magma oceans, where a tiny fraction of their volume almost instantaneously reaches the large exsolution depth associated to 10 $\rm M_{EO}$ (lower left corner in Figure~\ref{fig:Xexsolved_compa_TP_SE_01Meo_VS_10Meo}b).
Conversely, smaller water contents decrease the exsolution depth (Figure~\ref{fig:solubility_laws}) and slow down degassing, regardless of the planet size (Figures~\ref{fig:Xexsolved_compa_TP_SE_01Meo_VS_10Meo}a and \ref{fig:Xexsolved_compa_TP_SE_01Meo_VS_10Meo}c). For magma oceans with such a low water content, the time required to exsolve 100\% of oversaturated water is generally longer than the time spent in the fully molten stage (Figures~\ref{fig:Xexsolved_compa_TP_SE_01Meo_VS_10Meo}a and \ref{fig:Xexsolved_compa_TP_SE_01Meo_VS_10Meo}c), implying that water degassing is more likely to be incomplete or may not even have started before the end of the fully molten stage.

The fact that early degassing of water may be incomplete or even nonexistent in some cases has strong implications regarding the entire cooling history of the magma ocean, the resulting planetary surface conditions, and even the chemical differentiation and subsequent long-term evolution of the planet \citep{Maurice.etal2017,Ballmer.etal2017}. Indeed, if degassing is not efficient, the weak thermal blanketing effect of the resulting tenuous steam atmosphere will allow for efficient heat escape to space and rapid magma ocean cooling, making  degassing less efficient. Such a feedback would favor water storage in the mantle, resulting in early dry surface conditions, after rapid magma ocean solidification, with a water-rich solid-state mantle. This may have important implications regarding its properties and long-term evolution, including dynamics, convective regime, melting and degassing \citep[\eg][]{Lange1994, RegenauerLieb2001, Demouchy2016, Ni2016, Ohtani2020, Miyazaki2021, Miyazaki2022}. Conversely, a large initial concentration of water would induce an efficient and rapid exsolution of the magma ocean oversaturated water, which would, in turn, slow down the magma ocean cooling because of the efficient greenhouse effect of a thick steam atmosphere overlying the molten surface, and therefore leave time for the magma ocean to fully outgas its oversaturated  volatile content during the slow solidification process. The slow solidification of a magma ocean due to the presence of its thick steam atmosphere and the resulting loss of water through atmospheric escape has been proposed, as a consequence of the orbital distance, to explain the divergence between the Earth and Venus \citep{Hamano2013}. In addition, the atmospheric water content affects the spectral properties of exoplanets  \citep{Robinson2020, Smith2020} and high water contents atmospheres even likely experience a runaway greenhouse radius inflation that changes their mass-radius relationship in a possibly future detectable manner \citep{Turbet2019, Turbet2020}. These aspects emphasize the crucial role that magma ocean outgassing efficiency may play in planetary evolution, habitability, and exoplanets detectability and characterization.

\subsubsection{Model limitations and future directions}
By highlighting the influence of the dynamics of the fully molten magma ocean stage on its early degassing, our results motivate further investigations to characterize magma ocean degassing efficiency during its entire lifetime. This implies extending the approach presented here to values of Prandtl numbers larger than one, to better cover the different transient stages experienced throughout magma ocean solidification. One should note that due to the possible differences in convective dynamics in 2D and 3D simulations at $Pr < 1$ \citep{Schmalzl2004, Breuer2004}, the validity of our results may be limited to Prandlt numbers of unity and larger. Nevertheless, such values are relevant for terrestrial magma oceans \citep[\eg][and references therein]{Hoink2006, Solomatov2015, Maas2015, Maas2019}.
In addition, even though we accounted for the effect of rotation on the magnitude of convective velocity in a simplified way in our planetary outgassing model (Equation~\ref{eq:U_rotation}), the influence of planetary rotation in 3D geometry \citep{Maas2019} should also be explicitly considered in the fluid dynamics experiments.
Furthermore, our computational fluid dynamics experiments assumed homogeneous density and viscosity. Yet, these fluid properties may be strongly influenced by the volatile content \citep[\eg][]{Lange1994, Nikolaou.etal2019}, possibly leading to dynamic feedback mechanisms that may affect the efficiency of the  magma ocean to exsolve oversaturated volatiles, which  we could not directly evaluate in the presented numerical experiments. Such influences can be implemented since local changes in density and viscosity due to the volatile content can be numerically accounted for.   
Conducting a systematic exploration of this aforementioned governing parameter space will allow for a complete characterization of the exsolution efficiency (\ie in the form of the time evolution of the $d_{\rm min}$ distributions) of magma oceans at various stages of their evolution.

While such a task is out of the scope of the present paper, our numerical experiments and outgassing predictions show that for a significant range of magma ocean governing parameters, the degassing is not as efficient as previously thought \citep[localized and minimal to no degassing scenarios proposed in][]{Ikoma2018}. Therefore, the assumption of instantaneous outgassing of oversaturated volatile species may not always be valid and should only be considered with care.

Since the cooling rate of the magma ocean is controlled by its degassing via the amount of outgassed atmospheric greenhouse gases, our results suggest that magma ocean simulations shall consider the influence of magma ocean dynamics to self-consistently infer the magma ocean lifetime and the associated surface conditions.

The expressions we derived to predict the spectrum of the minimum depth/pressure reached by the convective fluid with time can be used in coupled atmosphere--mantle models. This may reveal additional complexities during and after the magma ocean evolution, due to the influence of volatile content on several key physical quantities that govern magma ocean and the subsequent solidified mantle convective dynamics, such as viscosity or density \citep[\eg][]{Lange1994, Litasov2007, Karki2010, Nikolaou.etal2019}.
Ultimately, this coupled approach will provide a considerably more accurate picture of the outgassing history of terrestrial magma oceans together with the formation of planetary atmospheres.

\section{Conclusions}
We investigated the influence of magma ocean convective dynamics on their degassing efficiency using numerical fluid dynamics modeling of thermal convection in 2D and 3D geometry at Prandtl number of unity and for various values of the Rayleigh number up to $10^{12}$. We measured and characterized the distribution of the minimum depth reached by the convecting fluid parcels at all times.
We found that the main characteristics of these distributions (average value and standard deviation) can simply be expressed as a function of the number of convective transits achieved, regardless of the value of the Rayleigh number.
This indicates that the ability of the magma ocean to exsolve volatiles is mainly governed by the magnitude of the convective velocities (or the effective Reynolds number), which is a function of the Rayleigh number considered. 
This simple relationship between convective velocity and outgassing efficiency allows deriving simple expressions that reproduce the results of our numerical experiments. 

We used our experimentally-derived expressions to estimate the magma ocean volume fraction that reaches any given exsolution depth as a function of time.
We found that a substantial number of convective transits need to be achieved for the entire vigorously convecting magma ocean volume to reach the exsolution depth, and thus, for all oversaturated volatile species to be exsolved out of the melt, form gas bubbles and degas at the surface. This is particularly important for volatile species that are highly soluble in the melt, such as water.
Depending on the value of the governing parameters considered, the time required to fully exsolve oversaturated volatile species can be comparable or even longer than the time spent in a fully molten transient stage. Therefore, in some cases, a fully molten mantle can start to crystallize before its entire volume has reached the exsolution depth. For example, despite vigorous convection, we found that complete outgassing of water can require more than $10^4$ years in a fully molten global terrestrial magma ocean, a duration that may be one order of magnitude larger than the lifetime of this earliest magma ocean transient stage. Consequently, volatile species remaining in the melt should be accounted for accordingly in the mass balance distributing them between the different planetary reservoirs during subsequent magma ocean evolution.
Degassing is generally less efficient in smaller planets due to their less vigorous convection that reduces the magnitude of convective velocities and exsolution rates. Conversely, on large planets, the convective vigor results in higher mixing rates that offset the additional transit times required by shallow exsolution depths.

Furthermore, if the convective vigor is low, the degassing is less efficient, which results in rapid cooling rates and short solidification timescales (due to the weak thermal blanketing effect of the tenuous outgassed atmosphere), which, in turn decreases outgassing efficiency by reducing the time available for the entire magma ocean to reach shallow depths. On the contrary, if the convective vigor is high, such as on large planets, the efficient outgassing will produce a thick and thermally opaque atmosphere, preventing efficient cooling of the magma ocean, and increasing its degassing efficiency since volatiles will have enough time to be fully exsolved during the slow solidification process.
Thus, the magma ocean convective dynamics, through the control of the outgassed atmosphere and resulting cooling rate may lead to divergent planetary evolution paths.

Finally, we show that degassing is strongly sensitive to the initial volatile content and that it is less efficient on planets with low water contents. This positive feedback between the initial water content and the degassing efficiency can in turn also lead to significantly different magma ocean evolutions: planets with low water contents would tend to poorly outgas and solidify quickly, rapidly leading to dry surface conditions, tenuous atmosphere and wet interiors propitious to melting, long-term tectonic and volcanic activity. In contrast, magma oceans with high initial water contents would lead to enhanced degassing, favoring a slow solidification under a thick and thermally opaque atmosphere, responsible for hot and extreme surface conditions, preventing the formation of water oceans and propitious to hydrodynamic escape that would desiccate the evolving planet. On the other hand, moderate, Earth-like water contents would imply moderate outgassing efficiency and intermediate solidification rates that are more adequate for temperate climates and stable liquid water at the planetary surface, with moderately dry interiors.
Therefore, the influence of magma ocean dynamics on its outgassing efficiency can significantly affect its subsequent thermal and chemical evolution.

Even though further investigations considering explicitly broader ranges of Rayleigh and Prandtl numbers, the  influences of volatiles on rheology and density, or rotation are required to better characterize the full degassing sequence of solidifying magma oceans, our results  demonstrate  that contrary to what is commonly assumed, even in a vigorously convecting magma ocean, degassing can be far from instantaneous or efficient.

%% file: appendix.tex
\appendix

\setcounter{figure}{0}
\setcounter{table}{0}

\section{Effect of solubility laws}  \label{solubility_laws}
\begin{figure}
\centering
\includegraphics[width=1.\textwidth, height=.5\textheight, keepaspectratio]{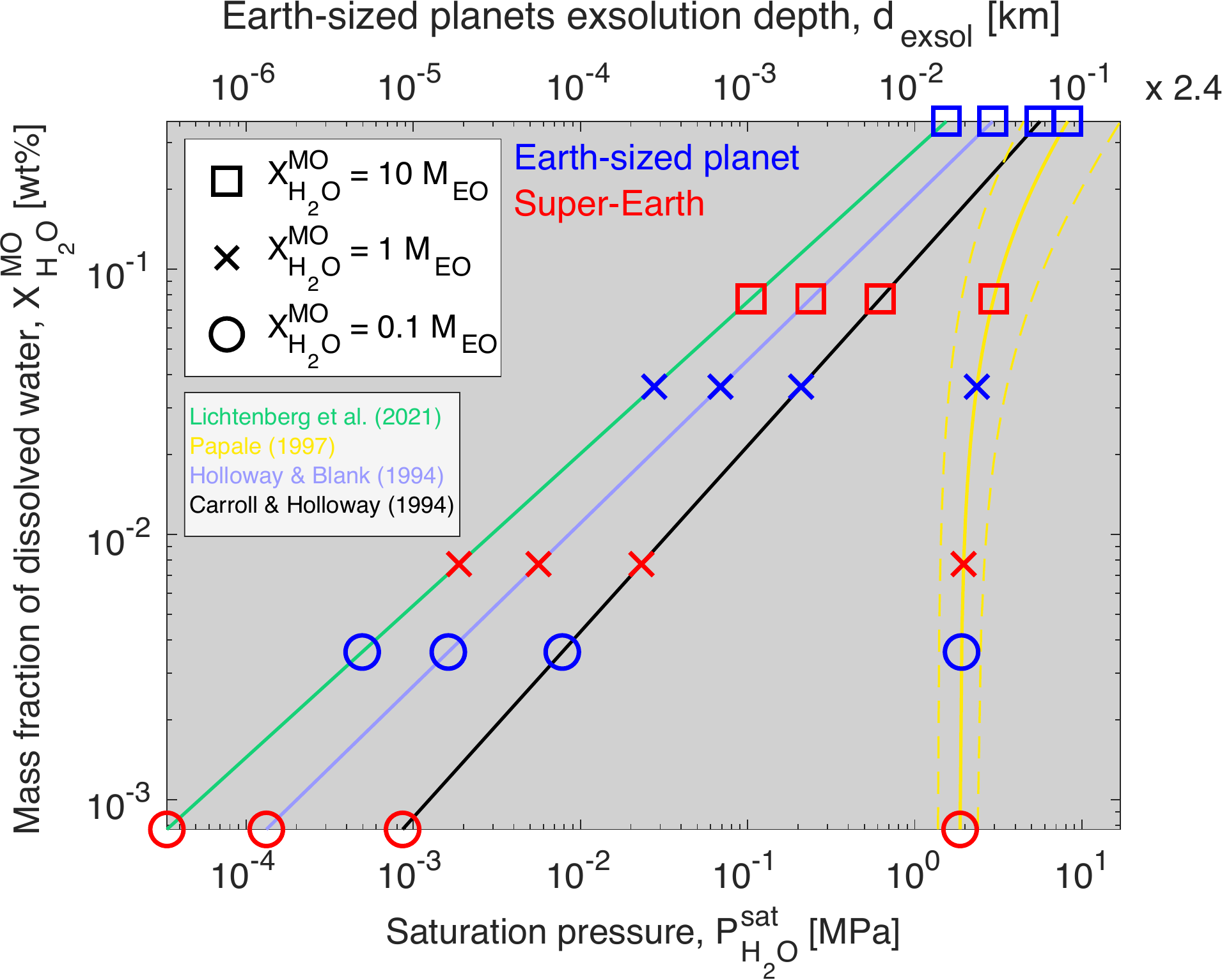}
\caption{H\textsubscript{2}O solubility laws in silicate melts. The saturation curve gives the location of the exsolution depth, $d_{\rm exsol}$, reported on Figure~\ref{fig:degassing_sketch} (blue dashed line). At a given cooling transient stage, an ascending fluid undergoes decompression, \ie a decrease of ambient lithostatic pressure, and goes from the right hand side to the left hand side of the diagram. As solidification proceeds, the silicate melt volume decreases and water partitions readily into the melt such that evolving magma ocean liquids are enriched in water, and the mass fraction of dissolved water $X_{\rm H_2O}^{\rm MO}$ increases along the saturation curve, from the bottom to the top of the diagram. In this study, we used the \cite{Carroll1994} solubility law (black, parameters given in Table~\ref{tab: solub_law}). Parameters of the other solubility laws are given in Table~\ref{tab:compa_solub}.}
\label{fig:solubility_laws}
\end{figure}

Solubility laws provide the saturation pressure, $P^{\rm sat}_{\rm vol}$, of any given volatile species as a function of its concentration in the melt. Thus, they give the overburden lithostatic pressure where a given volatile species has reached its maximum concentration in the melt and below which no more can be dissolved into the melt and start forming gas bubbles. Assuming hydrostatic equilibrium, the corresponding exsolution depth, $d_{\rm exsol}$, is obtained from Equation~\eqref{eq:hydrostat_eq}.
Figure~\ref{fig:solubility_laws} shows different solubility laws for water in silicate melts (see Table~\ref{tab:compa_solub} for the corresponding solubility laws coefficients).
In this study, we considered the \cite{Carroll1994} solubility law for water (Equation~\eqref{eq:solub_law}, Table~\ref{tab: solub_law}, black line in Figure~\ref{fig:solubility_laws}), which has been used in several magma ocean modeling studies \citep[\eg][]{Lebrun2013,Salvador2017,Nikolaou.etal2019,Bower2019} and gives intermediate saturation pressures compared to others. Solubility laws obtained for basaltic systems are generally more appropriate for magma oceans as basalts, \ie relatively low silica content melts, have high temperatures, low viscosities and high gas diffusivities \citep[\eg][]{Sparks1994,Lesher2015}. However, they tend to have more modest volatile contents than more viscous, silicic melts such as rhyolitic or dacitic melts. 
However, note that the viscosity of mafic and ultramafic melts (such as basalt and komatiite, respectively), is comparatively much less affected by the dissolved water content than that of polymerized, low temperatures, rhyolitic or dacitic melts \citep[\eg][their Figure 5.5 and references therein]{Lesher2015}.
To test the sensitivity of our model results to the solubility law considered, we plotted additional solubility laws obtained for basalts (light blue and yellow curves in Figure~\ref{fig:solubility_laws}, fitted from Figures 10C and A.1b in \cite{Holloway1994, Papale1997}, respectively), and derived from rhyolitic melts (green line in Figure~\ref{fig:solubility_laws}, from Table 1 in \cite{Lichtenberg.etal2021} and references therein).
For each solubility law considered, symbols mark the saturation pressure for three different water contents spanning three orders of magnitude, from 0.1 (circles) to 10 Earth ocean mass (squares), with one Earth ocean mass (1 $M_{\rm EO}$) equal to $1.4\times 10^{21}$ kg. Note that for the same mass of water, the mass fraction of dissolved water, $X^{\rm MO}_{\rm H_2O}$, which is the absolute mass of water relative to the mass of the melt, is larger for Earth-sized planets than for super-Earths as the same amount of water is dissolved and diluted within a larger, more massive, molten mantle in the latter. Considering such similar water contents for planets of different sizes implicitly makes the assumption that water endowment does not scale with planetary mass during the accretion sequence. Consequently this does not account for the possible effect of planet size on volatile abundance and assumes that water delivery is relatively less efficient on larger planets. This accretion scenario could be in agreement with the stochastic nature of volatile delivery, which may then be weakly dependent on planetary mass \citep[\eg][]{Raymond2006a,Raymond2007}. However, the uncertainties associated to volatile delivery sources and timing do not allow for a conclusive statement yet \citep[\eg][]{Marty2012, Morbidelli2012, OBrien2018, Venturini2020}.
As seen in Figure~\ref{fig:solubility_laws}, for a given water content, the solubility law chosen strongly influences the saturation pressure and thus the exsolution depth at which gas bubbles start to form. Conversely, the amount of water that can be dissolved into and outgassed out of the melt strongly varies depending on the solubility law chosen. These differences become more and more important when decreasing the mass fraction of dissolved water and are thus ($i$) larger for low water contents, and ($ii$) always larger for super-Earths than for Earth-sized planets for a given dissolved water mass.
As a result, on super-Earths, the saturation pressure of 0.1 Earth ocean mass can vary over more than four orders of magnitude solely depending on the arbitrary solubility law choice (red circles).
As shown in this study, the exsolution depth, and thus the choice of the solubility law, play a fundamental role regarding magma ocean degassing efficiency as the latter is determined by comparing the time required for the entire molten volume to reach this very depth and the time spent in a given transient stage (of given convective dynamics).
The solubility law considered is thus of fundamental importance when constraining the timing of atmospheric formation and magma ocean solidification, as well as the surface conditions reached at the end of this early thermal evolution stage.

\begin{figure}
\centering
\includegraphics[width=1.\textwidth, height=1\textheight, keepaspectratio]{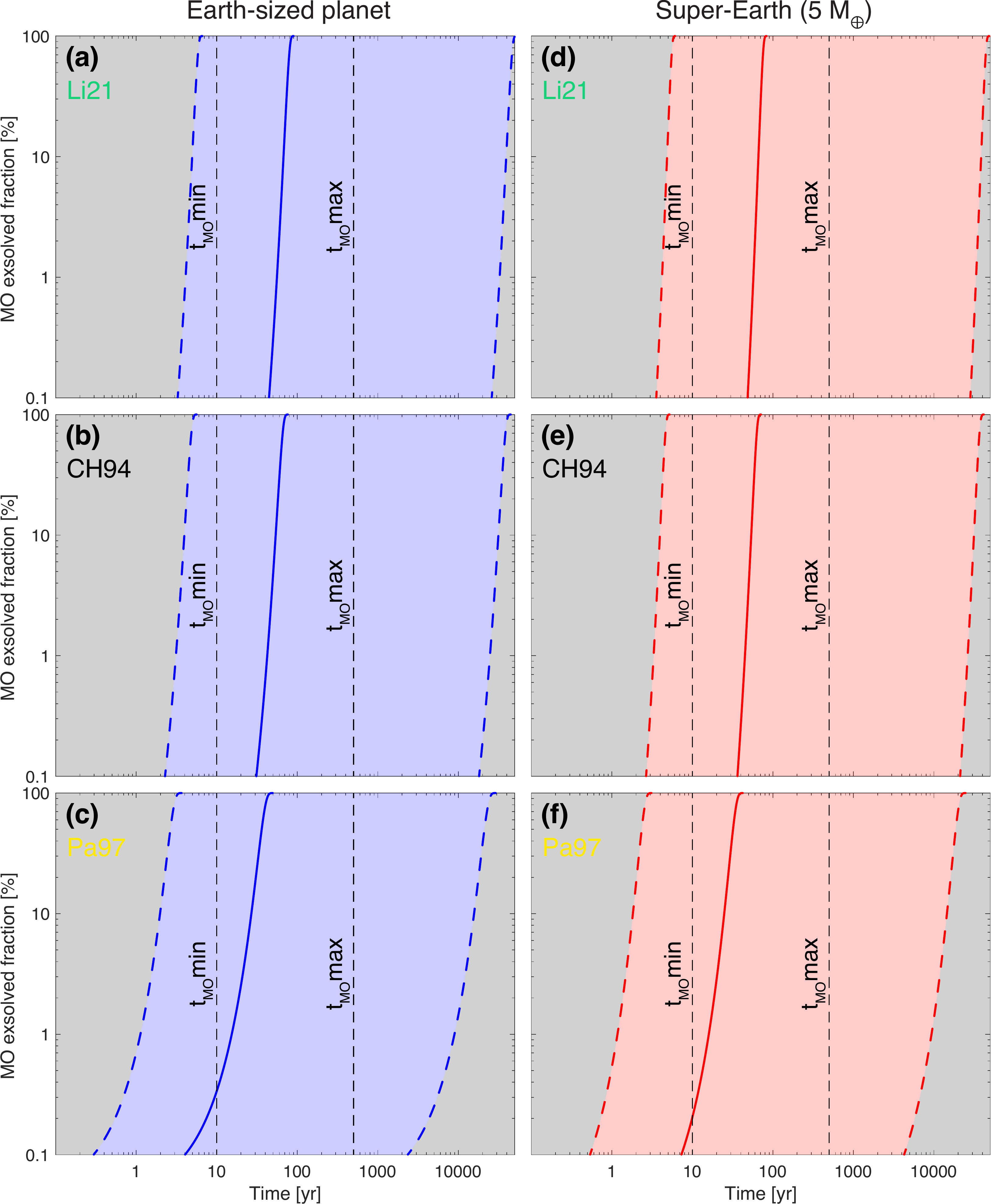}
\caption{Plausible range of the magma ocean exsolved fraction, \ie that reached the exsolution depth, as a function of time for different solubility laws, on (a-c) an Earth-sized planet and (d-f) a 5 Earth mass super-Earth, when considering 0.1 Earth ocean of water dissolved into the magma ocean. Solubility laws used to compute the exsolution depths are taken from (a,d) \cite{Lichtenberg.etal2021} (Li21), (b,e) \cite{Carroll1994} (CH94), and (c,f) \cite{Papale1997} (Pa97), and are shown in Figure~\ref{fig:solubility_laws} using the same color code (see Tables~\ref{tab: solub_law} and \ref{tab:compa_solub} for solubility laws parameters). Reference cases (plain lines) parameters for the Earth-sized planet and super-Earth are listed in Tables~\ref{tab: MO_parameters} and \ref{tab: MO_parameters_SE}, respectively. Parameters considered for the enhanced and weak degassing limits (blue and red dashed lines) are listed in Table~\ref{tab: degassing_cases_parameters}. Vertical black dashed lines indicate the minimum and maximum duration of the magma ocean in the fully molten transient stage from \cite{Lebrun2013} coupled magma ocean--atmosphere model.}
\label{fig:compa_solub_law_0.1Meo}
\end{figure}

\begin{figure}
\centering
\includegraphics[width=1.\textwidth, height=1\textheight, keepaspectratio]{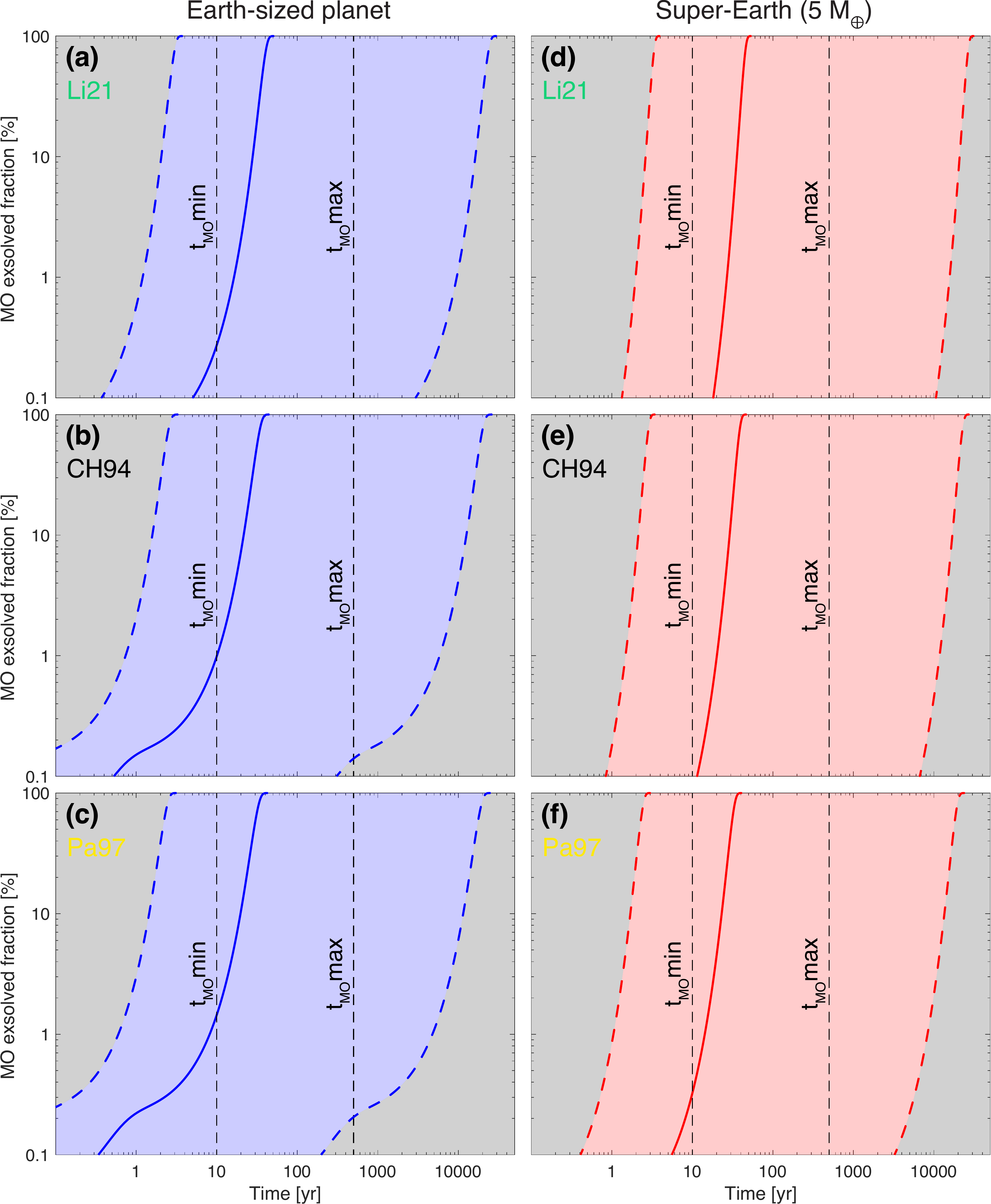}
\caption{Plausible range of the magma ocean exsolved fraction, \ie that reached the exsolution depth, as a function of time for different solubility laws, on (a-c) an Earth-sized planet and (d-f) a 5 Earth mass super-Earth, when considering 10 Earth oceans of dissolved water. Solubility laws used to compute the exsolution depths are taken from (a,d) \cite{Lichtenberg.etal2021} (Li21), (b,e) \cite{Carroll1994} (CH94), and (c,f) \cite{Papale1997} (Pa97), and are shown in Figure~\ref{fig:solubility_laws} using the same color code (see Tables~\ref{tab: solub_law} and \ref{tab:compa_solub} for solubility laws parameters). Reference cases (plain lines) parameters for the Earth-sized planet and super-Earth are given in Tables~\ref{tab: MO_parameters} and \ref{tab: MO_parameters_SE}, respectively. Parameters considered for the enhanced and weak degassing limits (blue and red dashed lines) are given in Table~\ref{tab: degassing_cases_parameters}. Vertical black dashed lines indicate  the minimum and maximum duration of the magma ocean in the totally molten transient stage from \cite{Lebrun2013} coupled magma ocean--atmosphere model.}
\label{fig:compa_solub_law_10Meo}
\end{figure}

Figures~\ref{fig:compa_solub_law_0.1Meo} and \ref{fig:compa_solub_law_10Meo} show how these other solubility laws would affect the timing of magma ocean oversaturated water exsolution for 0.1 and 10 $M_{\rm EO}$, respectively.
As emphasized above, the influence of the solubility law on degassing efficiency is always more important on super-Earths than on Earth-sized planets and low water contents are more sensitive to the solubility law choice. 
Finally, as discussed in the main text, one can note that regardless of the water content, degassing starts always earlier on Earth-sized planets due to their larger water concentrations and their larger associated exsolution depths compared to super-Earths (compare Figure~\ref{fig:compa_solub_law_0.1Meo}b with Figure~\ref{fig:compa_solub_law_0.1Meo}e and Figure~\ref{fig:compa_solub_law_0.1Meo}c with Figure~\ref{fig:compa_solub_law_0.1Meo}f). This effect is always more pronounced for the largest exsolution depths such as for 10 $M_{\rm EO}$ or with \cite{Papale1997} solubility law (compare Figure~\ref{fig:compa_solub_law_10Meo}b with Figure~\ref{fig:compa_solub_law_10Meo}e and Figure~\ref{fig:compa_solub_law_10Meo}c with Figure~\ref{fig:compa_solub_law_10Meo}f).
Yet, despite an earlier degassing start on Earth-sized planets, the more vigorous convection on super-Earths overcomes the effect of their shallower exsolution depths. This leads to faster and therefore more efficient outgassing on super-Earths. This leads to similar times to fully outgas oversaturated volatiles for both planet types for 10 $M_{\rm EO}$ (Figures~\ref{fig:compa_solub_law_10Meo}c and \ref{fig:compa_solub_law_10Meo}f) and is even responsible for slightly earlier complete exsolution times for super-Earths with 1 $M_{\rm EO}$ and 0.1 $M_{\rm EO}$, whose shallow exsolution depths make convective vigor the primary governing parameter for degassing efficiency (Figures~\ref{fig:Xexsolved_compa_TP_VS_SE_1Meo}, \ref{fig:compa_solub_law_0.1Meo}c and \ref{fig:compa_solub_law_0.1Meo}f), while the exact timing is influenced by the solubility law considered.

\begin{table}
\centering
\begin{threeparttable}
\caption{Parameters of the solubility laws shown in Figure~\ref{fig:solubility_laws} and to be used in Equation~\eqref{eq:solub_law}}
\label{tab:compa_solub}
\begin{tabular}{l c c}
\hline
Parameter & Value & Reference \\
\hline
Exponent $\beta_{\rm H_2O}$  & 1/0.61428                                  & Fitted from Figure 10C in \citet{Holloway1994} \\
Constant $\zeta_{\rm H_2O}$  & $0.045314\, {\rm wt\%~bar}^{-1}$  & Fitted from Figure 10C in \citet{Holloway1994} \\
Exponent $\beta_{\rm H_2O}$  & 1/0.67404                                  & Fitted from Figure A.1b in \citet{Papale1997}$^{*}$ \\
Constant $\zeta_{\rm H_2O}$  & $14.483\, {\rm wt\%~GPa}^{-1}$  & Fitted from Figure A.1b in \citet{Papale1997}$^{*}$ \\
Exponent $\beta_{\rm H_2O}$  & 1.747                                   & Table 1 in \citet{Lichtenberg.etal2021} \\
Constant $\zeta_{\rm H_2O}$  & $1.033\times10^{0}\, {\rm ppm~Pa}^{-1}$ & Table 1 in \citet{Lichtenberg.etal2021}\\
\hline
\multicolumn{3}{l}{$^{*}$Solubility law of the form: $P^{\rm sat}_{\rm H_2O}=((X^{\rm MO}_{\rm H_2O}+0.20882)/\zeta_{\rm H_2O})^{\beta_{\rm H_2O}}$.}
\end{tabular}
\end{threeparttable}
\end{table}

\section{Influence of the exsolution depth on the magma ocean exsolution time} \label{dexsol_influence}
\begin{figure}
\centering
\includegraphics[width=1.2\textwidth, height=1\textheight, keepaspectratio]{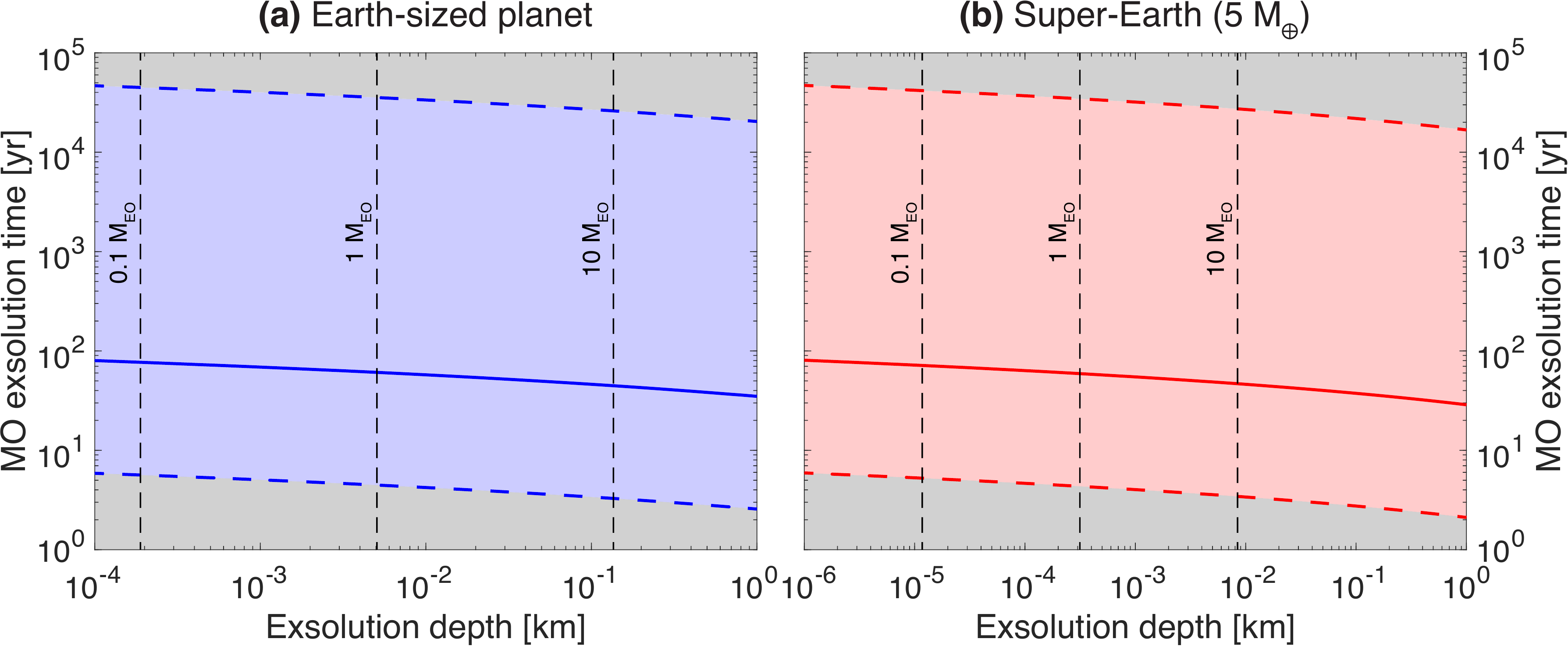}
\caption{Plausible range of time required for 99.99 \% of the magma ocean to reach the exsolution depth as a function of the exsolution depth for (a) an Earth-sized, and (b) a 5 Earth mass planet. Vertical dashed lines indicate the exsolution depths corresponding to 0.1, 1, and 10 Earth oceans of dissolved water. The values of the governing parameters used for the Earth-sized and super-Earth planets are listed in Tables~\ref{tab: MO_parameters} and \ref{tab: MO_parameters_SE}, respectively. These parameters have been used to compute the reference cases exsolution times (plain lines), and parameters for the weak (upper dashed lines), and enhanced (lower dashed lines) degassing limits are listed in Table~\ref{tab: degassing_cases_parameters}.}
\label{fig:exsol_time_VS_exsol_depth}
\end{figure}

One can estimate the efficiency of magma ocean outgassing using the exsolution time, $t_{\rm exsol}$, defined as the time required for 99.99\% of the magma ocean to reach a given exsolution depth at which its oversaturated water is fully exsolved and outgassed. This quantity is displayed as a function of the exsolution depth in Figure~\ref{fig:exsol_time_VS_exsol_depth} for the Earth-like and super-Earth reference cases (blue and red plain lines, whose parameters are given in Tables~\ref{tab: MO_parameters} and \ref{tab: MO_parameters_SE}, respectively), with the corresponding plausible ranges included between the weak and efficient degassing limits (blue and red areas in Figures~\ref{fig:exsol_time_VS_exsol_depth}a and \ref{fig:exsol_time_VS_exsol_depth}b, respectively, whose parameters are listed in Table~\ref{tab: degassing_cases_parameters}). As discussed above, an increase of the exsolution depth can result either from an increase of the initial concentration of dissolved water within a magma ocean of given size, or from the water incompatibility and resulting enrichment of the residual melt during magma ocean crystallization.
Regardless of planetary size, the timing for complete  magma ocean exsolution decreases when increasing the exsolution depth (Figures~\ref{fig:exsol_time_VS_exsol_depth}a and \ref{fig:exsol_time_VS_exsol_depth}b), since deep exsolution pressures are more rapidly reached, implying that magma ocean degassing is more efficient for larger water contents, as discussed above. Yet, one cannot state that magma ocean degassing becomes more and more efficient as crystallization proceeds only by looking at the exsolution depths and based on the fact that melt becomes more water-concentrated. Indeed, the decrease of the convection vigor with solidification would buffer the effect of larger exsolution depths. This underlines the importance of considering these two competing effects to accurately characterize magma ocean degassing efficiency.
Finally, depending on the duration of the fully molten stage, almost all scenarios in the parameters space explored predict that the magma ocean complete exsolution time is larger than $t_{\rm MO} {\rm min}=10$ years, thus generally preventing complete magma ocean degassing for this transient stage. Even when considering slow magma ocean cooling associated with the longest duration of the fully molten stage ($t_{\rm MO} {\rm max}=500$ years), about half of the scenarios, excluding the reference case, lead to incomplete or no degassing ($t_{\rm exsol} > t_{\rm MO}$).
In such cases, the amount of water that remained dissolved within the melt (because part or none of the latter have reached the exsolution depth) must be accounted for when considering the entire magma ocean cooling sequence.


\section{Benchmark results of the {\tt StreamV} code in 3D geometry and in the finite Prandtl number context} 
\label{3Dbenchmark}
We considered a convection benchmark proposed in \citet{Fusegi.etal91a}, which is an extension of the 2D test proposed in \citet{Davis.etal83} to 3D geometry.  It consists in a cubic domain in which two opposite vertical sidewalls (chosen here to be located perpendicular to the $x$-coordinate, while the $z$-axis is vertical and oriented opposite to the gravity vector) have homogeneous distinct temperatures. The four other walls are insulating, and no-slip conditions are imposed on all boundaries.
The material properties are homogeneous, and in this case the system is governed by the Prandtl number and the Rayleigh number already defined (with $Ra$ being defined here in terms of the temperature difference between the two isothermal vertical sidewalls).
The Prandtl number is set to 0.71, but the Rayleigh number is varied between $10^3$ and $10^6$. Convective currents progressively develop and the system is evolved until reaching steady state at which several quantities are compared:
(i) the average Nusselt number taken at the isothermal boundaries. In the conservative finite-volume formulation, the latter is the same on both isothermal boundaries at the steady sate;
(ii) the maximum values for the $x$ and $z$ components of the velocity field: $v_x^{\rm MP}{_{\rm max}}=\max [v_x(y=0.5)]$, $v_z^{\rm MP} {_{\rm max}}=\max [v_z(y=0.5)]$ in the symmetry plane ($y$=0.5);
(iii) the maximum values for each component of the velocity field in the entire domain: $v_{x_{\rm max}}$, $v_{y_{\rm max}}$, $v_{z_{\rm max}}$.
We performed a comparison with the study by \citet{Labrosse.etal2000} that documents all the diagnostics mentioned above. 
Cases for which $Ra$ is lower than $10^5$ are computed on a grid using $64 \times 64 \times 64$ cubic cells, while the other cases use $128 \times 128 \times 128$ cubic cells.
The results displayed in Table~\ref{tab:3DBM} show that the agreement is good with a relative error smaller than one percent for each quantities for each case considered.
\begin{table}
    \caption{Results of the 3D convection test between two vertical plates. Comparison of the results obtained with {\tt StreamV} on a regular grid with the results obtained in \citet{Labrosse.etal2000}  on their finest grid for each case considered}
    \label{tab:3DBM}
\vspace{1cm}    
\begin{tabular}{c|c|c|c}
Quantity & This study  &   \citet{Labrosse.etal2000} & Relative error in percents \\
\hline
\multicolumn{4}{c}{$Ra=10^3$} \\
\hline
 $Nu$                & 1.0715    &  1.0700   & 0.14 \\
$\max [v_x(y=0.5)]$  &   3.5360  &  3.54356  &  0.21 \\
$\max [v_z(y=0.5)]$  &   3.5365  &  3.54477  & 0.23   \\
$v_{x_{\rm max}}$    &   3.5360  & 3.54356   & 0.21 \\
$v_{y_{\rm max}}$    &   0.1723  &  0.17331  & 0.58 \\
$v_{z_{\rm max}}$    &   3.5365  &  3.54469  & 0.23 \\
\hline
\multicolumn{4}{c}{$Ra=10^4$} \\
\hline
 $Nu$                & 2.0604    &    2.0542 & 0.30  \\
$\max [v_x(y=0.5)]$  & 16.693    &  16.71986 & 0.16 \\
$\max [v_z(y=0.5)]$  & 18.619    &  18.6825  & 0.34 \\
$v_{x_{\rm max}}$    &  16.6928  &  16.7199  & 0.16 \\
$v_{y_{\rm max}}$    & 2.1401    &  2.15657  & 0.76 \\
$v_{z_{\rm max}}$    & 18.8847   &  18.9836  & 0.52 \\
\hline
\multicolumn{4}{c}{$Ra=10^5$} \\
\hline
 $Nu$                & 4.3446   &   4.3370 & 0.18\\
$\max [v_x(y=0.5)]$  & 43.1249  &  43.0610 & 0.15\\
$\max [v_z(y=0.5)]$  & 65.2327  &  65.4362 & 0.31\\
$v_{x_{\rm max}}$    & 43.9553  &  43.9037 & 0.12\\
$v_{y_{\rm max}}$    &  9.7256  &   9.6973 & 0.29 \\
$v_{z_{\rm max}}$    & 70.9028  &  71.0680 & 0.23 \\
\hline
\multicolumn{4}{c}{$Ra=10^6$} \\
\hline
 $Nu$                &   8.6985  &   8.6407   & 0.67   \\
$\max [v_x(y=0.5)]$  &  123.3032 &  123.4777  &  0.14 \\
$\max [v_z(y=0.5)]$  &  216.2094 &  218.2578  &  0.94 \\
$v_{x_{\rm max}}$    & 126.7283  &  126.9731  & 0.19  \\
$v_{y_{\rm max}}$    &  25.3554  &  25.5650   & 0.82 \\
$v_{z_{\rm max}}$    & 234.9607  &  236.7203  & 0.74 \\
\hline
\end{tabular}
\end{table}

%% file: highlights.tex
Highlights
\begin{itemize}
    \item Magma ocean dynamics governs volatile outgassing efficiency
    \item Volatile outgassing efficiency mainly depends on the magnitude of convective velocities
    \item The degassing efficiency decreases with planet size or initial water content
    \item Convective dynamics and cooling rate feedback may lead to divergent evolution paths
    \item Volatile outgassing in a vigorously convecting magma ocean may be considerably weaker than previously thought
\end{itemize}